\documentclass[preprint,nofootinbib,aps,superscriptaddress,tightenlines,floatfix]{revtex4}

\usepackage{graphicx}
\usepackage{epsfig}
\usepackage{dcolumn}

\newif\ifpdf
\ifx\pdfoutput\undefined
\pdffalse 
\else
\pdfoutput=1 
\pdftrue
\fi

\def\OMIT#1{{}}
\def\lqcd{\Lambda_{\rm QCD}}
\def\MSbar{\ensuremath{\overline{\rm MS}}}
\def\GeV{\mbox{GeV}}
\def\MeV{\mbox{MeV}}

\def\mBbar{\overline{m}_B}
\def\mDbar{\overline{m}_D}
\def\mbups{\ensuremath{m_b^{1S}}}
\def\mbps{\ensuremath{m_b^{\rm PS}}}

\def\mbms{\ensuremath{\overline{m}_b(m_b)}}

\def\lbar{\ensuremath{\bar\Lambda}}
\def\lups{\ensuremath{\Lambda^{1S}}}
\def\lps{\ensuremath{\Lambda^{\rm PS}}}
\def\lms{\ensuremath{\Lambda^{\MSbar}}}

\def\epsBLM{\ensuremath{\epsilon^2_{\rm BLM}}}
\def\d{{\rm d}}
\def\abs#1{ \left| #1 \right| }

\def\l#1{\ensuremath{\lambda_{#1}}}
\def\r#1{\ensuremath{\rho_{#1}}}
\def\t#1{\ensuremath{{\cal T}_{#1}}}

\newcommand{\nn}{\nonumber}
\newcommand{\beq}{\begin{equation}}
\newcommand{\eeq}{\end{equation}}
\newcommand{\beqa}{\begin{eqnarray}}
\newcommand{\eeqa}{\end{eqnarray}}

\begin{document}
\ifpdf
\DeclareGraphicsExtensions{.pdf, .jpg}
\else
\DeclareGraphicsExtensions{.eps, .jpg}
\fi

\preprint{ \vbox{\hbox{UCSD/PTH 02-19} \hbox{LBNL-51450} \hbox{UTPT 02-11} 
  \hbox{hep-ph/0210027} }}

\title{\boldmath $B$ decay shape variables and the precision determination\\
of  $|V_{cb}|$ and $m_b$\vspace*{8pt}}

\vspace*{1.5cm}

\author{Christian W.\ Bauer}\email{bauer@physics.ucsd.edu}
\affiliation{Department of Physics, University of California at San Diego,
  La Jolla, CA 92093\vspace{4pt} }

\author{Zoltan Ligeti}\email{zligeti@lbl.gov}
\affiliation{Ernest Orlando Lawrence Berkeley National Laboratory,
  University of California, Berkeley, CA 94720\vspace{4pt} }

\author{Michael Luke}\email{luke@physics.utoronto.ca}
\affiliation{Department of Physics, University of Toronto, 
  60 St.\ George Street, Toronto, Ontario, Canada M5S 1A7
  \\[20pt] $\phantom{}$ }

\author{Aneesh V.\ Manohar\vspace*{8pt}}\email{amanohar@ucsd.edu}
\affiliation{Department of Physics, University of California at San Diego,
  La Jolla, CA 92093\vspace{4pt} }

\begin{abstract} \vspace*{8pt}

We present expressions for shape variables of $B$ decay distributions in
several different mass schemes, to order $\alpha_s^2\beta_0$ and
$\lqcd^3/m_b^3$.  Such observables are sensitive  to the $b$ quark mass and
matrix elements in the heavy quark effective theory, and recent measurements
allow precision determinations of some of these parameters.  
We perform a combined fit to recent
experimental results from CLEO, BABAR, and DELPHI, and discuss the theoretical
uncertainties due to nonperturbative and perturbative effects. We discuss the
possible discrepancy between the OPE prediction, recent BABAR results and the
measured branching fraction to $D$ and $D^*$ states. We find $|V_{cb}|  =
\left( 40.8 \pm 0.9 \right) \times 10^{-3}$ and $m_b^{1S} = 4.74 \pm
0.10\,\text{GeV}$, where the errors are dominated by experimental uncertainties.
\end{abstract}

\maketitle

\section{Introduction}

The study of flavor physics and $CP$ violation is entering a phase when one is
searching for small deviations from the standard model. Therefore it becomes
important to revisit the theoretical predictions for inclusive decay rates and
their uncertainties, which provide clean ways to determine fundamental standard
model parameters and test the consistency of the theory.

Experimental studies of inclusive semileptonic and rare $B$ decays provide
measurements of fundamental parameters of the standard model, such as the CKM
elements $|V_{cb}|$, $|V_{ub}|$, and the bottom and charm quark masses.
Inclusive and rare decays are also sensitive to possible new physics
contributions, and the theoretical computations are model independent. The
operator product expansion (OPE) shows that in the $m_b \gg \lqcd$ limit 
inclusive $B$ decay rates are equal to the $b$ quark decay rates
\cite{OPE,book}, and the corrections are suppressed by powers of $\alpha_s$ and
$\lqcd/m_b$. High-precision comparison of theory and experiment requires a
precise determination of the heavy quark masses, as well as the matrix elements
$\lambda_{1,2}$, which parameterize the nonperturbative corrections to inclusive
observables at ${\cal O}(\lqcd/m_b)^2$. At order $(\lqcd/m_b)^3$, six new matrix
elements occur, usually denoted by $\r{1,2}$ and $\t{1,2,3,4}$. There are two
constrains on these six matrix elements, which reduces the number of parameters
that affect $B$ decays at order $(\lqcd/m_b)^3$ to four.

The accuracy of the OPE predictions depends primarily on the error of the quark
masses, and to a lesser extent on the matrix elements of these higher
dimensional operators.  It was proposed that these quantities can be determined
by studying shapes of $B$ decay spectra~\cite{volo,kl,FLSmass1,gremmetal}. 
Such studies have been recently carried out by the CLEO, BABAR and DELPHI
collaborations
\cite{cleohadron,cleophoton,dch,marina,Babarhadron,DELPHIlepton,DELPHIhadron}. 
A potential source of uncertainty in the OPE predictions is the size of
possible violations of quark-hadron duality~\cite{NIdual}.  Studying the shapes
of inclusive $B$ decay distributions may be the best way to constrain these
effects experimentally, since it should influence the relationship between
shape variables of different spectra.  Thus, testing our understanding of these
spectra is important to assess the reliability of the inclusive determination
of $\abs{V_{cb}}$, and also of $\abs{V_{ub}}$.

In this paper we present expressions for lepton and hadronic invariant mass
moments for the inclusive decay $B \to X_c \ell \bar\nu$, as well as photon
energy moments in $B \to X_s \gamma$ decays. We give these results as a
function of cuts on the lepton and photon energy, respectively.  Most results
in the  literature have been given in terms of the pole mass, which introduces 
artificially large perturbative corrections in intermediate steps,  making it
difficult to estimate perturbative uncertainties.  We present all results in
four different mass schemes: the pole mass, the $1S$ mass, the PS mass, and the
$\overline {\rm MS}$ mass.  We then carry out a combined fit to all currently
available data and investigate in detail the uncertainties on the extracted
parameters $|V_{cb}|$ and $m_b$.

The results of this paper can be combined with independent determinations of
the $b$ and $c$ quark masses from studies of $Q\bar Q$ states. We have chosen
not to discuss those constraints here, since there exist many detailed analyses
in the literature~\cite{masspaper}.  Furthermore, the determinations of $m_b$
and $m_c$ from $Q\bar Q$ states have theoretical uncertainties which are
totally different from the current extraction.  Consistency between the
extractions is therefore a powerful check on both determinations.

\section{Shape variables}

We study three different distributions, the charged lepton
spectrum~\cite{volo,gremmetal,GK,GS} and the hadronic invariant mass
spectrum~\cite{FLSmass1,FLSmass2,GK} in semileptonic $B\to X_c\ell\bar\nu$
decays, and the photon spectrum in $B\to X_s\gamma$~\cite{FLS,kl,llmw,bauer}. 
Similar studies are also possible in $B\to X_s\ell^+\ell^-$ and $B\to
X_s\nu\bar\nu$ decay~\cite{raremoments}, but at the present, such processes do
not give competitive information.

The $B\to X_c\ell\bar\nu$ decay rate is known to order $\alpha_s^2
\beta_0$~\cite{LSW} and $\lqcd^3 / m_b^3$~\cite{GK},  where $\beta_0=11-2n_f/3$
is the coefficient of the first term in the QCD $\beta$-function, and the terms
proportional to it often dominate at order $\alpha_s^2$.  For the charged
lepton spectrum we define shape variables which are moments of the lepton
energy spectrum with a lepton energy cut,
\begin{eqnarray}\label{Rdef}
R_0(E_0,E_1) = {\displaystyle \int_{E_1}
  {\d\Gamma\over \d E_\ell}\, \d E_\ell \over \displaystyle
  \int_{E_0} {\d\Gamma\over{\rm d}E_\ell}\, {\rm d}E_\ell }\,, \qquad
R_n(E_0) = {\displaystyle \int_{E_0} E_\ell^n\,
  {\d\Gamma\over{\rm d}E_\ell}\, \d E_\ell \over \displaystyle
  \int_{E_0} {\d\Gamma\over{\rm d}E_\ell}\, {\rm d}E_\ell }\,,
\end{eqnarray}
where $\d\Gamma/\d E_\ell$ is the charged lepton spectrum in the $B$ rest frame.
$R_n$ has dimension $\GeV^n$, and is known to order $\alpha_s^2
\beta_0$~\cite{GS} and $\lqcd^3 / m_b^3$~\cite{GK}.  Note that these
definitions differ slightly from those in Ref.~\cite{gremmetal}, and follow the
CLEO~\cite{dch,marina} notation. The DELPHI collaboration~\cite{DELPHIlepton}
measures the mean lepton energy and its variance (both without any energy cut),
which are equal to $R_1(0)$ and $R_2(0)-R_1(0)^2$, respectively.

For the $B\to X_c\ell\bar\nu$ hadronic invariant mass spectrum we define the
mean hadron invariant mass and its variance, both with \emph{lepton} energy
cuts $E_0$,
\beq\label{Sdef}
S_1(E_0) = \langle m_X^2 - \mDbar^2 \rangle \Big|_{E_\ell>E_0}, \qquad
S_2(E_0) = \Big\langle (m_X^2 - \langle m_X^2\rangle)^2 \Big\rangle
  \Big|_{E_\ell>E_0},
\eeq
where $\mDbar = (m_D+3m_{D^*})/4$ is the spin averaged $D$ meson mass.  It is
conventional to subtract $\mDbar^2$ in the definition of the first moment
$S_1(E_0)$. $S_n$ has dimension (GeV)$^{2n}$, and is known to order $\alpha_s^2
\beta_0$~\cite{FLSmass2} and $\lqcd^3 / m_b^3$~\cite{GK}.  For a given $E_0$,
the maximal kinematically allowed hadronic invariant mass is $m_X^{\rm max} =
\sqrt{m_B^2 - 2m_B E_0}$.  Once $m_X^{\rm max} - \mDbar \gg \lqcd$, the OPE is
expected to describe the data.  

The above shape variables can be combined in numerous ways to obtain
observables that may be more suitable for experimental studies because of
reduced correlations.  For example, $S_1$ and $R_0$ can be combined to obtain
predictions for
\beq\label{neat}
\langle m_X^2 -\mDbar^2 \rangle \Big|_{E_1 > E_\ell>E_0}
= {S_1(E_0) R_0(0,E_0) - S_1(E_1) R_0(0,E_1) \over 
  R_0(0,E_0) - R_0(0,E_1)} \,,
\eeq
that allows comparing regions of phase space that do not overlap~\cite{Oliver}.

For $B\to X_s\gamma$, we define the mean photon energy and variance, with a
photon energy cut $E_0$,
\beq\label{Tdef}
T_1(E_0) = \langle E_\gamma \rangle \Big|_{E_\gamma>E_0}, \qquad
T_2(E_0) = \Big\langle (E_\gamma - \langle E_\gamma\rangle)^2 \Big\rangle
  \Big|_{E_\gamma>E_0},
\eeq
where $\d\Gamma/\d E_\gamma$ is the photon spectrum in the $B$ rest frame.
Again, $T_{1,2}$ are known to order $\alpha_s^2 \beta_0$~\cite{llmw} and
$\lqcd^3 / m_b^3$~\cite{bauer}.  In this case the OPE is expected to describe
$T_i(E_0)$ once $m_B/2-E_0 \gg \lqcd$. Precisely how low $E_0$ has to be to
trust the results can only be decided by studying the data as a function of
$E_0$; one may expect that $E_0 = 2\,\GeV$ available at present is sufficient.
Note that the perturbative corrections included are sensitive to the
$m_c$-dependence of the $b\to c\bar c s$ four-quark operator ($O_2$)
contribution.  This is a particularly large effect in the $O_2-O_7$
interference~\cite{llmw}, but its relative influence on the moments of the
spectrum is less severe than that on the total decay rate. The variance, $T_2$,
is very sensitive to any boost of the decaying $B$ meson; this contribution
enhances $T_2$ by $\beta^2/3$ at leading order~\cite{llmw}, where $\beta$ is
the boost ($\beta \simeq 0.064$ if the $B$ originates from $\Upsilon(4S)$
decay).  This is absent if $\d \Gamma/\d E_\gamma$ is reconstructed from a
measurement of $\d \Gamma/\d E_{m_{X_s}}$.

\section{Mass schemes}

The OPE results for the differential and total decay rates are given in terms
of the $b$ quark mass, $m_b$, and the quark mass ratio, $m_c/m_b$.  (Throughout
this paper quark masses without other labels refer to the pole mass.)  The pole
mass can be related to the known meson masses via the $1/m_Q$ expansion
\beq\label{mesonmass}
m_M = m_Q + \lbar - \frac{\l1+d_M\l2(m_Q)}{2m_Q}
  + \frac{\r1+d_M\r2}{4m_Q^2} - \frac{\t1+\t3+d_M(\t2+\t4)}{4m_Q^2} + \ldots\,,
\end{equation}
where $m_M$ ($M=P,V$) is the hadron mass, $m_Q$ is the heavy quark mass, and
$d_P=3$ for pseudoscalar and $d_V=-1$ for vector mesons.  The $\lambda_i$'s and
$\rho_i$'s  are matrix elements of local dimension-5 and 6 operators in HQET,
respectively, while the $\t{i}$'s are matrix elements of time ordered products
of operators with terms in the HQET Lagrangian, and are defined
in~\cite{GK}\footnote{These are related to the parameters $\rho^3_D$,
$\rho^3_{LS}$, $\rho_{\pi\pi}^3$, $\rho_{\pi G}^3$, $\rho_S^3$ and $\rho_A^3$ 
introduced in \cite{BSUV95}.}. The ellipses denote $\lqcd^4/m_Q^3$ corrections,
which can be neglected to the order we are working.  Using
Eq.~(\ref{mesonmass}), we can eliminate $m_c$ in favor of $m_b$ and the higher
order matrix elements,
\beq\label{massdiff}
m_b - m_c = \mBbar - \mDbar 
  - \l1 \left( \frac{1}{2m_c} - \frac{1}{2m_b} \right)
  + (\r1 - \t1 - \t3) \left( \frac{1}{4m_c^2} - \frac{1}{4m_b^2} \right) ,
\eeq
where $\overline m_M = (m_P+3m_V)/4$ denotes the spin averaged meson masses.

Only three linear combinations of $\t{1-4}$ appear in the expressions for $B$
meson decays (a fourth linear combination would be required to describe $B^*$
decays).  The reason is that the $\t{1-4}$ terms originate from two sources: (i)
the mass relations in Eqs.~(\ref{mesonmass}) and (\ref{massdiff}) which depend
on $\t1+\t3$ and $\t2+\t4$; and (ii) corrections to the order $\lqcd^2/m_b^2$
terms in the OPE, which amount to the replacement $\l1 \to \l1 + (\t1 +
3\t2)/m_b$ and $\l2 \to \l2 + (\t3+3\t4)/(3m_b)$.  Since $\t1 + 3\t2 = (\t1+\t3)
+ 3(\t2+\t4) - (\t3+3\t4)$, only three linear combinations are independent.
Therefore, we may set $\t4=0$, and the fit then  projects on the linear
combinations
\beq\label{crazycombinations}
\t1 - 3\t4\,, \qquad \t2 + \t4\,, \qquad \t3 + 3\t4\,.
\eeq

The mass splittings between the vector and pseudoscalar mesons,
\begin{equation}
\Delta m_M \equiv m_V - m_P = \frac{2\,\kappa(m_Q)\, \l2(m_b)}{m_Q}
  - \frac{\r2-(\t2+\t4)}{m_Q^2} + \ldots\,,
\end{equation}
constrain the numerical values of some of the HQET matrix elements.
Here $\kappa(m_c) = \big[ \alpha_s(m_c)/\alpha_s(m_b) \big]^{3/\beta_0} \sim
1.2$ is the scaling of the magnetic moment operator between $m_b$ and $m_c$.
In terms of the measured $B^*-B$ and $D^*-D$ mass splittings, 
$\Delta m_B$ and $\Delta m_D$,
\begin{eqnarray}
\l2(m_b) &=& { m_b^2\, \Delta m_B - m_c^2\, \Delta m_D \over
  2[m_b-\kappa(m_c)\,m_c]}\,, \label{m3const1}\\
\r2-(\t2+\t4) &=& {m_b m_c [\kappa(m_c)\, m_b\, \Delta m_B - m_c\, \Delta m_D]
  \over  m_b - \kappa(m_c)\,m_c}\,. \label{m3const2}
\end{eqnarray}
These equations differ slightly from those in Ref.~\cite{GK}, and are
consistent to order $1/m_Q^3$.  Since order $\alpha_s(\lqcd/m_Q)^2$ corrections
in the OPE have not been computed, whether we set $\kappa(m_c)$ to its physical
value, $\kappa(m_c) \simeq 1.2$, or to unity is a higher order effect that
cannot be consistently included at present.  Using $\kappa(m_c) = 1.2$ or 1 in
the fits gives effects which are negligible compared with other uncertainties
in the calculation.

It is well-known that the pole masses suffer from a renormalon
ambiguity~\cite{renormalon}, which only cancels in physical observables against
a similar ambiguity in the perturbative expansions~\cite{luke_renormalon}. 
Although any quark mass scheme can be used to relate physical observables to
one another, the neglected higher order terms may be smaller if a
renormalon-free scheme is used.  When using pole masses it is important to
always work to a consistent order in the perturbative expansion, since
$\bar\Lambda$ can have large changes at each order in perturbation theory, even
though the relations between measurable quantities such as the shape variables
and the total semileptonic decay rate have much smaller changes.  Since
$\bar\Lambda$ depends strongly on the order of the calculation in perturbation
theory, one can get a misleading impression about the convergence of the
calculation, and its uncertainties.  The advantage of using renormalon-free
mass schemes is that the convergence may be manifest.

Several mass definitions which do not suffer from this ambiguity have been
proposed in the literature, and we consider here the \MSbar, $1S$, and PS
masses.  (There is a renormalon ambiguity in the $1S$ and PS masses, but it is
of relative order $\lqcd^4/m_b^4$ and so is irrelevant for our
considerations.)  The \MSbar\ mass is related to the pole mass through
\beq\label{massMS}
\frac{\mbms}{m_b} = 1 - \epsilon\, \frac{\alpha_s(m_b)C_F}{\pi} 
  - \epsilon^2\, 1.562\, \frac{\alpha_s(m_b)^2}{\pi^2}\, \beta_0
  + \ldots \,.
\eeq
and $C_F = 4/3$ in QCD. The parameter $\epsilon \equiv 1$ is a new expansion
parameter, which for the \MSbar\ mass is the same as the order in $\alpha_s$.
While the \MSbar\ mass is appropriate for high energy
processes, such as $Z$ or
$h \to b\,\bar b$, it is less useful in processes where the typical momenta are
below $m_b$.  The \MSbar\ mass is defined in full QCD with dynamical $b$ quarks
and is appropriate for calculating the scale dependence above $m_b$.  However,
it does not make sense to run the \MSbar\ mass below $m_b$; this only
introduces spurious logarithms that have no physical significance.  Thus,
although the \MSbar\ mass is well-defined, it is not a particularly useful
quantity to describe $B$ decays.  Therefore, several ``threshold mass"
definitions have been introduced that are more appropriate for low energy
processes.  

The $1S$ mass is related to the pole mass through the perturbative
relation~\cite{ups1,ups2}
\beq\label{mass1S}
\frac{\mbups}{m_b} = 1 - \frac{[\alpha_s(\mu) C_F]^2}{8} \left[1 \epsilon
  + \epsilon^2\, \frac{\alpha_s(\mu)}{\pi} 
  \left(\ell+\frac{11}{6} \right) \beta_0 + \ldots \right] ,
\eeq
where the right hand side is the mass of the $\Upsilon(1S)$ $\bar b b$ bound
state as computed in perturbation theory, and $\ell= \ln[\mu/(\alpha_s(\mu)\,
C_F\, m_b)]$.  For the $1S$ mass there is a subtlety in the perturbative
expansion due to a mismatch between the order in $\epsilon$ and the order in
$\alpha_s$, so that terms of order $\alpha_s^{n+1}$ in Eq.~(\ref{mass1S}) are
of order $\epsilon^n$~\cite{ups1}. 

The potential-subtracted (PS) mass \cite{3loopPS} is defined with respect to a
factorization scale $\mu_f$.  It is related  to the pole mass through the
perturbative relation
\beq\label{massPS}
\frac{\mbps(\mu_f)}{m_b} = 1 
  - \frac{\alpha_s(\mu) C_F}{\pi}\, \frac{\mu_f}{m_b} \left[ 1\epsilon 
  + \epsilon^2\, \frac{\alpha_s(\mu)}{2 \pi} 
  \left(\ell + \frac{11}{6}\right) \beta_0 + \ldots \right] ,
\eeq
where now $\ell = \ln(\mu/\mu_f)$. In this paper we will choose $\mu_f=2\,\GeV$.

Another popular definition is the kinetic, or ``running", mass $m_b(\mu)$
introduced in~\cite{BSUV95,BSUV97}.  The kinetic mass has properties similar to
the PS mass, since it is defined with a cutoff that explicitly separates long-
and short-distance physics.  It should give comparable results, so we will not
consider it here.  We note, however, that in this scheme matrix elements such
as $\lambda_1$ are also naturally defined with respect to a momentum cutoff. 
This has the advantage of absorbing some ``universal" radiative corrections
into the definitions of the matrix elements instead of the coefficients in the
OPE, and is expected to improve the  behavior of the perturbative series
relating $\lambda_1$ to physical quantities.  However, as usual,  the
perturbative relation between physical quantities is unchanged, and adopting
this definition leaves our fits to $|V_{cb}|$ and $m_b$ unchanged. 

The results for the various shape variables are functions of the $b$ quark
mass.  To simplify the expressions, in analogy with $\bar\Lambda$ defined in
Eq.~(\ref{mesonmass}), we define new hadronic parameters by the following
relations
\beqa\label{leqn}
\lups &=& \frac{m_\Upsilon}{2} - \mbups\,, \nn\\
\lps &=& \frac{m_\Upsilon}{2} - \mbps\,, \nn\\
\lms  &=& 4.2\, {\rm GeV} - \mbms\,. 
\eeqa
We will refer to $\bar\Lambda$, $\lups$, $\lps$ and $\lms$ generically as
$\Lambda$. Note that the introduction of $\Lambda$ is purely for computational
convenience. The form Eq.~(\ref{leqn}) is chosen so that the value of $\Lambda$
is numerically of order $\lqcd$. We can therefore expand the radiative
corrections in powers of $\Lambda$ and keep only the leading term and the first
derivative. This is convenient because it avoids having to compute the
radiative corrections, which  involve a lengthy numerical integration, for each
trial value of the quark mass in the fit.  Note also that in the $1S$, PS and
$\MSbar$ schemes the dependence on $m_B-m_b$ is purely kinematic and is 
treated exactly, although it is formally of order $\lqcd$.

Thus, the decay rates will be expressed in terms of 9 parameters: the
$\Lambda$'s in each mass scheme which we treat as order $\lqcd$, two parameters
of order $\lqcd^2$, \l1\ and \l2, and six parameters of order $\lqcd^3$, \r1,
\r2, and $\t{1-4}$.  Of these, only 6 are independent unknowns, as \l2\ is
determined by Eq.~(\ref{m3const1}), $\r2-(\t2+\t4)$ is determined by
Eq.~(\ref{m3const2}), and $\t4$ can be set to zero as explained preceding
Eq.~(\ref{crazycombinations}).

\section{Expansions and their convergence}

The computations in this paper include contributions of order $1/m_Q^2$ and
$1/m_Q^3$, as well as radiative contribution of order $\epsilon$, and $\epsBLM$,
the so-called BLM contribution at order $\epsilon^2$ which is proportional to
$\beta_0$. 
The dominant theoretical errors arise from the higher order terms which we have
neglected. In the perturbative series, we have neglected the non-BLM part of the
two-loop correction. We have
also neglected the unknown order $\alpha_s/m_b^2$ and $1/m_b^4$ corrections in
the OPE.  The decay distributions depend on the charm quark mass, which is
determined from $\mBbar-\mDbar$ using Eq.~(\ref{massdiff}). This formula
introduces $\lqcd^4/m_c^4$ corrections. Since $m_c$ only enters inclusive decay
rates in the form $m_c^2/m_b^2$, the largest $1/m^4$ corrections are of order
$\lqcd^4/(m_b^2m_c^2)$. Finally, the $O(\epsilon\lbar)$ corrections for $S_1$
and $S_2$ have only been  calculated without a cut on the lepton energy
\cite{FLSmass2}.

For the $B\to X_c\ell\bar\nu$ decay rate and the shape variables defined in
Eqs.~(\ref{Rdef}), (\ref{Sdef}), and (\ref{Tdef}) we give results in the
Appendix in the four different mass schemes discussed, for the coefficients
$X^{(1-17)}(E_0)$  in the expansion
\beqa\label{expdef}
X(E_0) &=& X^{(1)}(E_0) + X^{(2)}(E_0)\, \Lambda 
  + X^{(3)}(E_0)\, \Lambda^2 + X^{(4)}(E_0)\, \Lambda^3 \nn\\*
&+& X^{(5)}(E_0)\, \l1 + X^{(6)}(E_0)\, \l2 
  + X^{(7)}(E_0)\, \l1 \Lambda + X^{(8)}(E_0)\, \l2 \Lambda \nn\\*
&+& X^{(9)}(E_0)\, \r1 + X^{(10)}(E_0)\, \r2 
  + X^{(11)}(E_0)\, \t1 + X^{(12)}(E_0)\, \t2 \\*
&+& X^{(13)}(E_0)\, \t3 + X^{(14)}(E_0)\, \t4 
  + X^{(15)}(E_0)\, \epsilon + X^{(16)}(E_0)\, \epsBLM
+ X^{(17)}(E_0)\, \epsilon \Lambda \,,\nn
\eeqa
where $X(E_0)$ is any of $R_0(0,E_0)$, $R_i(E_0)$, $S_i(E_0)$, or $T_i(E_0)$
and $i=1,2$.  Note that to obtain $R_0(E_0,E_1)$ one needs to reexpand
$R_0(0,E_1) / R_0(0,E_0)$, but using $R_0(0,E_0)$ allows us to tabulate the
results as a function of only one variable.  The expressions for $R_0(0,E_0)$
are also convenient for deriving the predictions for other observables, such as
those in Eq.~(\ref{neat}).

Unfortunately there is no simple way to relate the results in different mass
schemes, because a particular value of the physical $E_0$ cut corresponds to
different limits of integrations in the dimensionless variables (such as
$2E_0/m_b$) in different mass schemes.  We list the coefficients of the
expansions of the shape variables in the various mass schemes in the
Appendix.

Before using these expressions, one has to assess the convergence of both the
perturbative expansions and of the power suppressed corrections.  As each shape
variable arises from a ratio of two series, the result can be worse or better
behaved than the individual series in the numerator and denominator.  We have
checked that this is the reason for the apparent poor behavior of, for example,
$R_1(1.5\,{\rm GeV})$ in the $1S$ scheme, where one sees that order $\alpha_s$
term $R_1^{(15)}(1.5\,{\rm GeV}) = 0.001$, whereas the order $\alpha_s^2$ BLM
term $R_1^{(16)}(1.5\,{\rm GeV}) = 0.003$ is larger.  Since separately the
numerator and denominator show good convergence, one should not conclude that
$R_1(1.5\,{\rm GeV})$ is not a useful observable to constrain the HQET
parameters. In general, one cannot conclude whether a series is poorly behaved
or not by comparing the $\alpha_s^2$ term with the $\alpha_s$ term because of
possible cancellations. Instead, one should compare with the expected size of
terms based on a naive dimensional estimate.

In Refs.~\cite{FLSmass1,FLSmass2} the second hadronic invariant mass moment
defined as $\langle (m_X^2 -\mDbar^2)^2 \rangle$ was studied, and it was
observed that the size of the $\lqcd^3/m_b^3$ correction was comparable to both
the $\lqcd^2/m_b^2$ and $\alpha_s\lqcd/m_b$ terms.  The authors therefore
concluded that the convergence of the OPE was suspect for this moment, and 
argued that useful constraints on $\lbar$ and $\l1$ could not be obtained.  A
very similar situation holds for the variance $S_2$.  However, one can  obtain
more insight into the convergence of this moment by examining the behavior of
the relevant terms in the  OPE for $\langle m_X^2\rangle$ and $\langle
m_X^4\rangle$ separately.   In the pole scheme (for simplicity), the
expressions are
\begin{eqnarray}
{1\over m_B^2}\left.\langle m_X^2\rangle\right\vert_{E_\ell>0}
&=& {m_D^2\over m_B^2} + 0.24 \frac{\lbar}{m_B} + 0.26 \frac{\lbar^2}{m_B^2} 
  + 1.02 \frac{\l1}{m_B^2} + 2.2 \frac{\r1}{m_B^3} 
  + 0.21 \frac{\alpha_s}{4\pi} 
  + 0.41 \frac{\alpha_s}{4 \pi} \frac{\bar \Lambda}{m_B} \,, \nn\\
{1\over m_B^4}\left.\langle m_X^4\rangle\right\vert_{E_\ell>0}
&=& \frac{m_D^4}{m_B^4} + 0.07 \frac{\lbar}{m_B} + 0.14 \frac{\lbar^2}{m_B^2} 
  + 0.15 \frac{\l1}{m_B^2} - 0.23 \frac{\r1}{m_B^3} 
  + 0.08 \frac{\alpha_s}{4 \pi} 
  + 0.27 \frac{\alpha_s}{4\pi} \frac{\bar \Lambda}{\bar m_B} \,. \nn\\
\end{eqnarray}
The OPE for both observables is well behaved, with the canonical size of the
$\rho_1$ term a factor of 5--10 smaller than the $\lambda_1$ term.   The
corresponding constraints in the $\lbar-\l1$ plane have slopes which differ by
roughly a factor of two, and so constrain one linear combination of $\lbar$ and
$\l1$ much better than the orthogonal combination.

If instead of the second moment we consider the variance, we may combine the
two series to find 
\beq
{1\over m_B^4}\left.\langle m_X^4 - \langle
m_X^2\rangle^2\rangle\right\vert_{E_\ell>0} = 0.01 \frac{\bar \Lambda^2}{\bar
m_B^2}
- 0.14 \frac{\lambda_1}{\bar m_B^2} 
- 0.86 \frac{\rho_1}{\bar m_B^3} 
+ 0.02  \frac{\alpha_s}{4 \pi} 
+ 0.06 \frac{\alpha_s}{4 \pi} \frac{\bar \Lambda}{\bar m_B} \,.
\eeq
The variance gives constraints in the $\lbar-\l1$ plane which are almost
orthogonal to those of the first moment, but since it is simply a linear
combination of the first and second moments, it cannot constrain the parameters
any better. However, it is also no worse: none of the coefficients are larger
than would be expected by dimensional analysis.  The apparent poor convergence
of the variance is due to a cancellation in the $\lbar$ (and to a lesser extent
the $\l1$) terms between the two series.  Therefore, there is no reason to
expect the $O(1/m_B^4)$ terms to be anomalously large. Constraints arising from
$S_2$ (or from $\langle (m_X^2 -\mDbar^2)^2 \rangle$) therefore need not be
dismissed, although they are very sensitive to $\rho_1$ and so are of limited
utility unless a sufficiently large number of observables is measured that
$\rho_1$ is also constrained.

\section{Experimental Data}\label{sec:data}

The experimental data for the lepton spectrum from the CLEO collaboration are
the three lepton moments~\cite{dch,marina}
\beqa\label{cleolepton}
&& R_0(1.5\, \GeV,1.7\,\GeV) = 0.6187 \pm 0.0021, \nn\\
&& R_1(1.5\, \GeV) = (1.7810 \pm 0.0011)\, \GeV, \nn\\
&& R_2(1.5\, \GeV) = (3.1968 \pm 0.0026)\, \GeV^2.
\eeqa
For $R_0$ and $R_1$, we used the averaged electron and muon values, with the
full correlation matrix as given in Ref.~\cite{dch}. For $R_2$, we have used
the weighted average of the electron and muon data~\cite{marina}. The DELPHI
collaboration measures the lepton energy and variance~\cite{DELPHIlepton},
\beqa
&& R_1(0) = (1.383 \pm 0.015)\, \GeV, \nn\\
&& R_2(0)-R_1(0)^2 = (0.192 \pm 0.009)\, \GeV^2.
\eeqa

For the hadronic invariant mass spectrum we have CLEO measurements of the mean
invariant mass and variance with a lepton energy cut of
1.5\,GeV~\cite{cleohadron}
\beqa
  S_1(1.5\, \GeV) &=& (0.251 \pm 0.066)\, \GeV^2, \nn\\
  S_2(1.5\, \GeV) &=& (0.576 \pm 0.170)\, \GeV^4 ,
\eeqa
and DELPHI measurements of the mean invariant mass and variance with no lepton
energy cut~\cite{DELPHIhadron} 
\beqa
  S_1(0) &=& (0.553 \pm 0.088)\, \GeV^2, \nn\\
  S_2(0) &=& (1.26 \pm 0.23)\, \GeV^4. 
\eeqa
Both collaborations also measure the second moment $\langle (m_X^2 -\mDbar^2)^2
\rangle$, but we do not use this result since it is not independent of $S_1$
and $S_2$.

The BABAR collaboration measures the first moment of the hadron spectrum for
various values of the lepton energy cut~\cite{Babarhadron}. The data points are
highly correlated, and the variation of the first moment with the energy cut
appears to be in poor agreement with the OPE predictions. We will do our fits
without the BABAR data, as well as including the BABAR data for the two extreme
values of their lepton energy cut, $E=0.9$ and $E=1.5$\,GeV~\cite{Babarhadron},
to avoid overemphasizing many points with correlated errors in the fit,
\beqa
  S_1(1.5\, \GeV) &=& (0.354 \pm 0.080)\, \GeV^2, \nn\\
  S_1(0.9\, \GeV) &=& (0.694 \pm 0.114)\,  \GeV^2.
\eeqa
Note that we took into account that CLEO~\cite{dch} and
BABAR~\cite{Babarhadron} used $\mDbar = 1.975\,$GeV to obtain the quoted values
of $S_1$, whereas DELPHI~\cite{DELPHIhadron} used $\mDbar = 1.97375\,$GeV.

For the photon spectrum we use the CLEO results~\cite{cleophoton}
\begin{eqnarray}
T_1(2\, \GeV) &=& (2.346 \pm 0.034) \, \GeV, \nn\\
T_2(2\, \GeV) &=&  (0.0226 \pm 0.0069) \, \GeV^2.
\end{eqnarray}

The final piece of data is the semileptonic decay width, for which we use the
average of $B^\pm$ and $B^0$ data~\cite{pdg},
\beq\label{slrateexp}
\Gamma(B \to X \ell \bar \nu) = (42.7 \pm 1.4) \times 10^{-12}\, {\rm MeV}.
\eeq
We do not average this with the $B_s$ and $b$-baryon semileptonic widths, as
the power suppressed corrections can differ in these decays.

Eqs.~(\ref{cleolepton})--(\ref{slrateexp}) provide a total of 14 measurements
that enter our fit.

\section{The Fit}

In this section we perform a simultaneous fit to the various 
experimentally measured moments and the semileptonic rate.  It is 
important to note that we do not include any correlations between 
experimental measurements beyond those presented in \cite{dch,marina}, 
and so the experimental uncertainties are not completely taken into 
account.  Nevertheless, the fit demonstrates the importance of 
including the full correlation of the $O(1/m_b^3)$ terms in the 
different observables, and also indicates the relative importance of 
the theoretical and experimental uncertainties.

We use the fitting routine Minuit to fit simultaneously for the shape variables
and the total semileptonic branching fraction, by minimizing $\chi^2$, and 
present results for the fit in the $1S$ scheme (the other schemes give
comparable results).

In addition to the experimental uncertainties, there are also uncertainties in
the theory because the formulae used in the fit are not exact. From naive
dimensional analysis we find the fractional theory errors $0.0003$ from
$\left(\alpha_s/4\pi \right)^2$ terms,  $0.0002$ from  $\left(\alpha_s/4\pi
\right) \lqcd^2/m_b^2$ terms, and $0.001$  from $\lqcd^4/(m_b^2m_c^2)$ terms.
In some cases, naive dimensional analysis underestimates the
uncertainties, and an alternative is estimating the uncertainties by the size
of the last term computed in the perturbation series. We combine these
estimates by adding in quadrature half of the $\epsBLM$ term and a $0.001
m_B^n$ theoretical error for quantities with mass dimension $n$. In computing
$\chi^2$, we add this theoretical error in quadrature to the experimental
errors. This procedure avoids giving a large weight in the fit to a very
accurate measurement that cannot be computed reliably.  Because the
perturbative results in the $1S$ scheme are not expected to be artificially
badly behaved (as they are in, for example, the pole scheme) this estimate of
the perturbative uncertainty should be reasonable.  We will examine the
convergence of perturbation theory later in this section.

The unknown matrix elements of the $1/m_b^3$ operators are the largest source of
uncertainty in the fit.  One expects these matrix elements to be of order
$\lqcd^3$.  To allow for this theoretical input, we include an additional
contribution to $\chi^2$ from the matrix elements of each $1/m_b^3$ operators,
$\rho_{1,2}$ and $\t{1-4}$, that we denote generically by $\langle \mathcal{O}
\rangle$, 
\beq\label{delchi}
\Delta\chi^2(m_\chi,M_\chi) = \cases{
0\,,  &  $|\langle {\cal O} \rangle| \le m_\chi^3$\,, \cr
\left[|\langle \mathcal{O} \rangle| - m_\chi^3\right]^2 / M_\chi^6 \,,~~~
  &  $|\langle {\cal O} \rangle| > m_\chi^3$\,, \cr}
\eeq
where $(m_\chi,M_\chi)$ are both thought of as quantities of order $\lqcd$.
This way we do not prejudice $\langle \mathcal{O} \rangle$ to have any
particular value in the range $|\langle {\cal O} \rangle| \le m_\chi^3$.  In
the fit we take $M_\chi=500\,\MeV$, and vary $m_\chi$ between $500\,\MeV$ and 
$1\,\GeV$  to test that our results for $|V_{cb}|$ and $m_b$ are insensitive to
this input (our final results are obtained with $m_\chi=500\,\MeV$).  The data
are sufficient to constrain the $1/m_b^3$ operators in the sense that they can
be consistently fit with reasonable values, but they are not determined with
any useful precision.  Finally, since only three linear combinations of
$\t{1-4}$ appear in the formulae, we fit setting $\t4=0$, so that the fit
values for $\t{1-3}$ with this choice for $\t4$ are the values of $\t1-3\t4$,
$\t2+\t4$, and $\t3+3\t4$.

\begin{table}[t]
\caption{Fit results for $|V_{cb}|$, $m_b$, $\lambda_1$ and
$\lambda_1+(\t1+3\t2)/m_b$ in the $1S$ scheme. The $|V_{cb}|$ value includes
electromagnetic radiative corrections; see Eq.~(\ref{vcb}).   The upper/lower
blocks are fits excluding/including the BABAR data, and have 5 and 7 degrees of
freedom, respectively.\label{table1}}
\begin{tabular}{c||c|c|c|c|c}
$m_\chi\ [\GeV]$&  $\chi^2$ &  $|V_{cb}|\times 10^3$  &  $m_b^{1S}\, [{\rm
GeV}]$  &  $\lambda_1\, [{\rm GeV}^2]$ & $\lambda_1+{\t1+3\t2\over m_b}\, [{\rm
GeV}^2]$
\\
\hline\hline
$0.5$ &
$5.0$ &
$40.8\pm 0.9$ &
$4.74\pm 0.10$ &
$-0.22\pm 0.38$ &
$-0.31\pm 0.17$ 
\\
$1.0$ & 
$3.5$ & 
$41.1 \pm 0.9$ &
$4.74\pm 0.11$& 
$-0.40\pm 0.26$ &
$-0.31\pm 0.22$ 
\\
\hline
\hline
$0.5$ &
12.9 &
$40.8\pm 0.7$ &
$4.74\pm 0.10$ &
$-0.14\pm 0.13$ &
$-0.29\pm 0.10$ 
\\
$1.0$ & 
$8.5$ & 
 $40.9 \pm 0.8$ & 
$4.76 \pm 0.11$ & 
$-0.22 \pm 0.25$ &
$-0.17 \pm 0.21$ 
\\
\hline\hline
\end{tabular}
\end{table}
\begin{table}[t]
\caption{Fit results for the $1/m_b^3$ coefficients in the $1S$ scheme. The
upper/lower blocks are fits excluding/including  the BABAR data. The constraint
in Eq.~(\ref{m3const2}) is used to determine $\rho_2$. 
\label{tab:m3}}
\[
\begin{array}{c||c|c|c|c}
m_\chi\ [\GeV]&\rho_1 & \rho_2 & \t1+\t3 & \t1+3\t2 \\
\hline\hline
0.5& 
0.15 \pm 0.12 & 
-0.01 \pm 0.11 & 
-0.15 \pm 0.84 & 
-0.45 \pm 1.11 
\\
1.0& 
0.16 \pm 0.18 & 
-0.05 \pm 0.16 & 
0.41 \pm 0.40 & 
0.45 \pm 0.49
\\
\hline\hline
0.5& 
0.17 \pm 0.09 & 
-0.04 \pm 0.09 & 
-0.34 \pm 0.16 & 
-0.66 \pm 0.32 
\\
1.0& 
0.08 \pm 0.18 & 
-0.12 \pm 0.15 & 
0.11 \pm 0.33 & 
0.23 \pm 0.47 
\\
\hline\hline
\end{array}
\]
\end{table}

The fit results are summarized in Tables I and II. In Table~\ref{table1} we show
the results of the fit for $|V_{cb}|$, $\mbups$ and $\lambda_1$, as well as the
``effective" combination $\lambda_1+(\t1+3\t2)/m_b$ which enters in the OPE, and
which, due to correlated errors, is better constrained than $\lambda_1$.  From
these results we can also obtain an expression for $|V_{cb}|$ as a function of
the semileptonic branching ratio and the $B$ meson lifetime. We find
\begin{equation}\label{vcb}
|V_{cb}|  = \left( 40.8 \pm 0.7 \right) \times 10^{-3}\, \eta_{\text{QED} }
\left[ { \mathcal{B}(B \to X_c \ell \bar\nu) \over 0.105}
{1.6 \text{ps} \over \tau_B} \right]^{1/2} .
\end{equation}
The quoted error contains all uncertainties from $m_b$, $\lambda_1$, the
$1/m_b^3$ matrix elements, as well as perturbative uncertainties. The parameter
$\eta_{\text{QED}}\sim 1.007$ is the electromagnetic correction to the inclusive
decay rate, which has been included in the values for $|V_{cb}|$ presented in
Table~I. Including the BABAR data increases the $\chi^2$ by about a factor of
two. Doubling the allowed range of the $1/m_b^3$ parameters increases the
uncertainties only minimally and reduces $\chi^2$ somewhat.

\begin{figure}[t]
\includegraphics[width=10cm]{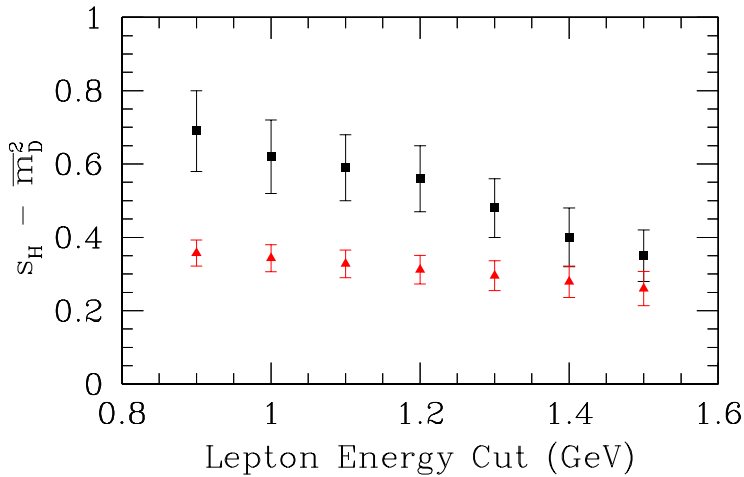}
\caption{Comparison of the BABAR measurement of the hadron invariant mass
spectrum \cite{Babarhadron} vs.\ the lepton energy cut (black squares), and our
prediction from the fit not including BABAR hadronic mass data (red triangles).
\label{fig:babarsh}}
\end{figure}

The reason we carried out separate fits excluding and including the BABAR data
on $S_1(E_0)$ is because of its inconsistency at low $E_0$ with the fit done
without it.  To see this, note that on very general grounds $S_1(E_0)$ is a
monotonically decreasing function of $E_0$.  The theoretical prediction
corresponding to the fit in the first line of Table~I is $S_1(0) = (0.42 \pm
0.03)\,\GeV^2$, which is significantly below the lowest BABAR data point,
$S_1(0.9\,\GeV) = (0.694 \pm 0.114)\,\GeV$.  Assuming that the branching ratio
to nonresonant channels between $D^*$ and $D^{**}$ is negligible, this
prediction for $S_1(0)$ implies an upper bound on the fraction of excited
(i.e., non-$D^{(*)}$) states in $B\to X_c\ell\bar\nu$ decay~\cite{GK}, which is
below 25\%, and is in contradiction with the measured $B\to D^{(*)}\ell\bar\nu$
branching fractions.  To resolve this, either the assumption  that low-mass
nonresonant channels are negligible could be wrong, or some measurements or 
the theory have to be several standard deviations off. The $X_c$ spectrum
effectively has this (assumed) feature in the CLEO and BABAR analyses but not
in DELPHI.  It is thus crucial to precisely and model independently measure the
$m_{X_c}$ distribution in semileptonic $B\to X_c\ell\bar\nu$ decay. A
comparison of the BABAR hadronic moment data with our fit is given in
Fig.~\ref{fig:babarsh}.

To get more insight into the obtained uncertainties, we have performed several
additional fits in which we turn off individual contributions to the errors.
Here we present the results for the fits with $m_\chi = 0.5$ and not including
the BABAR data. Similar results are true when the BABAR values are included.
Neglecting all $1/m_b^3$ terms, as well as the naive estimate of the theoretical
uncertainties gives a fit with $\chi^2=81$ for 9 degrees of freedom. Including
only the $1/m_b^3$ terms gives $\chi^2=21$ for 5 degrees of freedom. This is a
vastly better fit, reducing $\chi^2$ by about 60 by adding only 4 new
parameters. Nevertheless, the fact that $\chi^2$ per degree of freedom is about
5 shows that there is a statistically significant discrepancy between theory and
experiment if other theoretical uncertainties are not included. Only after
including this estimate do we get $\chi^2/{\rm dof} \approx 1$. We also
estimated the size of the theoretical uncertainties by setting all experimental
errors to zero. This reduces all uncertainties by roughly a factor of three.
Thus, the fit is dominated by experimental uncertainties.

The fit gives a value of the $b$ quark mass which is consistent with other
extractions, and with an uncertainty at the 100\,MeV level. For comparison,
$\Upsilon$ sum rules extractions in Refs.~\cite{hoang,benekesigner} give
$m_b^{1S}= 4.69 \pm 0.03\,\GeV$ and $m_b^{1S}= 4.78 \pm 0.11\,\GeV$,
respectively by a fit to the $\bar BB$ system near threshold. The error on
$\lambda_1$ is larger than previous extractions from $T_1$ and $S_1$
\cite{cleohadron}, because we are including more conservative estimates of the
theoretical uncertainties. Despite this, the uncertainty on $|V_{cb}|$ is
smaller than from previous extractions. Note that we have only used the value
of the semileptonic branching ratio of $B$ mesons.  It is inconsistent to
combine the average semileptonic branching ratio of $b$ quarks (including $B_s$
and $\Lambda_b$ states) with the moment analyses, since hadronic matrix
elements have different values in the $B/B^*$ system, and in the $B_s/B_s^*$ or
$\Lambda_b$.

The fit results for the $1/m_b^3$ parameters are shown in Table~\ref{tab:m3}. 
Clearly, one is not able to determine the values of the $1/m_b^3$ parameters
from the present fit. All that can be said is that the preferred values are
consistent with dimensional estimates. There is also some indication that
$\rho_2$ is small, as is expected in some models~\cite{GK}.

One can also use the fits to predict other observables that can be measured. For
example, we predict the values for the fractional moments $R_{3a}$,
$R_{3b}$, $R_{4a}$, $R_{4b}$, $D_3$ and $D_4$ given by Bauer and
Trott~\cite{BT}. The predicted values are given in Table~\ref{tab:bt}.
\begin{table}[t]
\caption{Fit predictions for fractional moments of the electron spectrum. The
upper/lower blocks are fits excluding/including the BABAR data.
\label{tab:bt}}
\[
\begin{array}{c||c|c|c|c|c|c}
m_\chi\ [\GeV]&R_{3a} & R_{3b} & R_{4a} & R_{4b} & D_3 & D_4 \\
\hline\hline
0.5&
0.302 \pm 0.003 &
2.261 \pm 0.013 &
2.127  \pm 0.013&
0.684 \pm 0.002 &
0.520 \pm 0.002 &
0.604 \pm 0.002\\
1.0&
0.302 \pm 0.002 &
2.261 \pm 0.011 &
2.128  \pm 0.011 &
0.684 \pm 0.002 &
0.519 \pm 0.002 &
0.604 \pm 0.001\\
\hline\hline
0.5&
0.302 \pm 0.002 &
2.261 \pm 0.012 &
2.127  \pm 0.012 &
0.684 \pm 0.002 &
0.520 \pm 0.002 &
0.604 \pm 0.001\\
1.0&
0.302 \pm 0.002 &
2.262 \pm 0.012 &
2.129  \pm 0.012 &
0.684 \pm 0.002 &
0.519 \pm 0.001 &
0.604 \pm 0.001\\
\hline\hline
\end{array}
\]
\end{table}
The results are robust, and do not depend on the width chosen for the $1/m^3$
operators, or whether or not we include the BABAR data.

\begin{figure}[t]
\includegraphics[width=10cm]{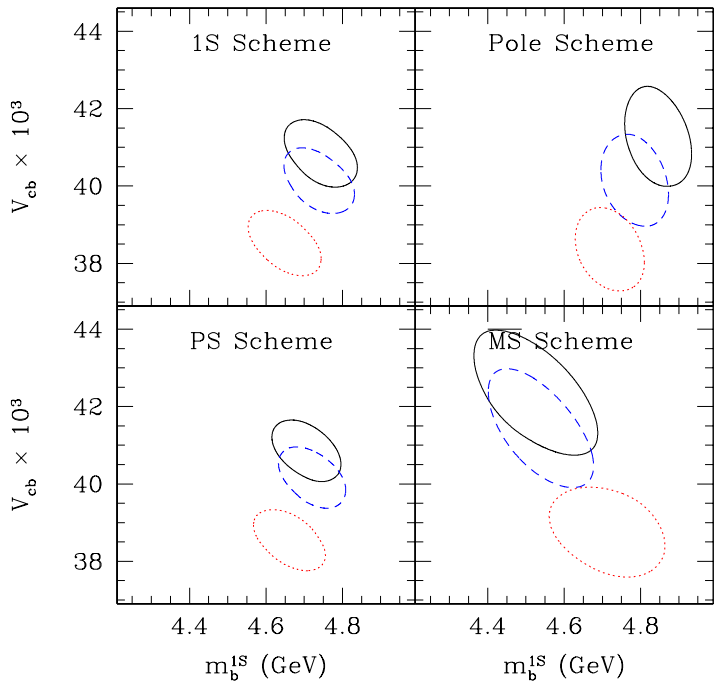}
\caption{The $1\sigma$ error ellipse in the $m_b^{1S}$ vs.\ $|V_{cb}|$ plane,
using different mass schemes for the fit.  For each scheme we show the contours
obtained at tree level (dotted red curves), at order $\epsilon$ (dashed blue
curves), and at order $\epsBLM$ (solid black curves).\label{fig2}}
\end{figure}

Finally, it is useful to study the convergence of perturbation theory by
carrying out the fit at different orders in the perturbation expansion.  In
Figure~\ref{fig2} we show the $1\sigma$ error ellipse in the $m_b^{1S}$ vs.\
$|V_{cb}|$ plane, for the four different mass schemes.  For each scheme we show
three contours, obtained at tree level (dotted red curves), at order $\epsilon$
(dashed blue curves), and including order $\epsBLM$ corrections as well (solid
black curves). For each of these curves, the conversion of the fitted mass to
the  ${1S}$ mass has been done at the consistent order in perturbation theory.
One can see that the convergence of the perturbative expansion is slightly
better for the $1S$ and the PS schemes compared with the pole scheme.   This is
because there is an incomplete cancellation of formally higher order terms,
such as $\alpha_s\lbar^2$, which are large in the pole scheme. The larger
uncertainties in the \MSbar\ scheme are due to large contributions at BLM
order, which are included in the uncertainty estimate, as explained at the
beginning of this section.

\section{Summary and conclusions}

Experimental studies of the shape variables discussed in this paper are crucial
in determining from experimental data the accuracy of the theoretical
predictions for inclusive $B$ decays rates, which rest on the assumption of
local duality.  Detailed knowledge of how well the OPE works in different
regions of phase space (and a precise value of $m_b$) will also be important
for the determination of $|V_{ub}|$ from inclusive $B$ decays. A serious
discrepancy between theory and data would imply, for example for $|V_{cb}|$,
that only its determination from exclusive decays has a chance of attaining
a reliable error below the $\sim 5\%$ level.

\begin{figure}[ht]
\includegraphics[width=0.5\textwidth]{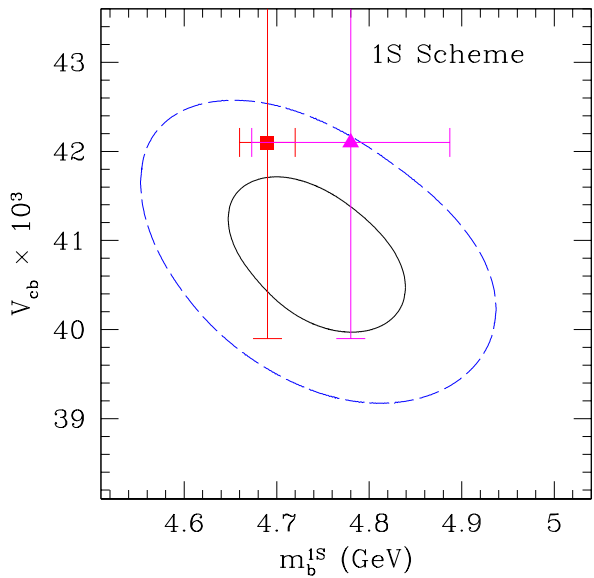}
\caption{The 1- and 2-$\sigma$ regions in the $m_b^{1S}$
vs.\ $|V_{cb}|$ plane using the $1S$ mass scheme.  Superimposed are the values
and errors of the determination of $|V_{cb}|$ from exclusive decays~\cite{pdg}
and that of $\mbups$ from sum rules in Ref.~\cite{hoang} (red square) and
Ref.~\cite{benekesigner} (magenta triangle).\label{fig:sum}}
\end{figure}

The analysis in this paper shows that at the present level of accuracy, the
data from the lepton and photon spectra are consistent with the theory, with no
evidence for any breakdown of quark-hadron duality in shape variables.  Two
related problems at present are the BABAR measurement of the average hadronic
invariant mass as a function of the lepton energy cut and the total branching
fraction to $D$ and $D^*$ states, both of which appear problematic to reconcile
with the other measurements combined with the OPE.  However, both problems
depend on assumptions about the invariant mass distribution of the decay
products, which needs to be better understood.  Excluding the BABAR data and
the problem of the  $B\to D^{(*)}\ell\bar\nu$ branching ratios, the fit 
provides a good description of the experimental results, with $\chi^2=5.0$ for
12 data points and 7 fit parameters in the $1S$ scheme.

The main results (in the $1S$ scheme) are summarized in Fig.~\ref{fig:sum}
where we compare our determination of $|V_{cb}|$ and $\mbups$ with those from
exclusive $B$ decays and upsilon sum rules.   We obtain the following values:
\begin{eqnarray}\label{finalresults}
|V_{cb}|  &=& \left( 40.8 \pm 0.9 \right) \times 10^{-3}, \nn \\
m_b^{1S} &=& (4.74 \pm 0.10)\,\text{GeV}. 
\end{eqnarray}
This corresponds to the $\overline{{\rm MS}}$ mass  $\overline m_b(\overline
m_b) = 4.22 \pm 0.09\,\text{GeV}$. We have also presented the value of
$|V_{cb}|$ as a function of the semileptonic branching ratio and the $B$ meson
lifetime
\begin{equation}
|V_{cb}|  = \left( 41.1 \pm 0.7 \right) \times 10^{-3}\,
\left[ { \mathcal{B}(B \to X_c \ell \bar\nu) \over 0.105}
{1.6 \text{ps} \over \tau_B} \right]^{1/2} .
\end{equation}
We have constrained the $1/m^3$ matrix elements and predicted the values for 
fractional moments of the electron spectrum to better than 1\% accuracy.

Setting experimental errors to zero gives errors in $|V_{cb}|$ and $m_b^{1S}$
of $0.35 \times 10^{-3}$ and $35$\,MeV, respectively.  These numbers indicate
the theoretical limitations, although their precise values depend on details of
how the theoretical uncertainties are estimated.  If the agreement between the
experimental results improve in the future, then a full two loop calculation of
the total semileptonic rate and of $B\to X_c\ell\bar\nu$ decay spectra would
help to further reduce the theoretical uncertainty in $|V_{cb}|$ and $m_b$.

\begin{acknowledgments}

We thank our friends at CLEO, BABAR and DELPHI for numerous discussions related
to this work.
C.W.B.\ thanks the LBL theory group and Z.L.\ thanks the LPT-Orsay for their
hospitality while some of this work was completed.
This work was supported in part by the US Department of Energy under contract
DE-FG03-97ER40546 (C.W.B.\ and A.V.M.); by the Director, Office of Science,
Office of  High Energy and Nuclear Physics, Division of High Energy Physics, of
the  U.S.\ Department of Energy under Contract DE-AC03-76SF00098 and by a DOE
Outstanding Junior Investigator award (Z.L.); and  by the Natural Sciences and
Engineering Research Council of Canada (M.L.).

\end{acknowledgments}

\appendix*
\section{Coefficient functions in various mass schemes}

\setcounter{topnumber}{5}
\renewcommand\topfraction{1}
\renewcommand\floatpagefraction{0}

In this Appendix we give numerical results for the the $B\to X_c\ell\bar\nu$
decay rate and the shape variables defined in Eqs.~(\ref{Rdef}), (\ref{Sdef}),
and (\ref{Tdef}), in the four mass schemes discussed.  For all quantities the
coefficients of the expansions are defined as in Eq.~(\ref{expdef}), and all
numerical values are in units of GeV to the appropriate power.  We use
$\alpha_s(m_b) = 0.22$ and the spin- and isospin-averaged meson masses,
$\mBbar=5.314\,\GeV$ and $\mDbar=1.973\,\GeV$.

\subsection{\boldmath The $1S$ mass scheme}

The $B\to X_c\ell\bar\nu$ decay width in the $1S$ scheme is given by
\beqa\label{upswidth}
\Gamma(B\to X_c\ell\bar\nu) &=& {G_F^2\, |V_{cb}|^2\over 192\pi^3}\,
  \bigg({m_\Upsilon\over2}\bigg)^5\, \bigg[ 0.534 
  -0.232\, \Lambda -0.023\, \Lambda^2 + 0.\, \Lambda^3 \nn\\*
&& -0.11\, \l1 -0.15\, \l2 
  -0.02\, \l1 \Lambda + 0.05\, \l2 \Lambda \nn\\*
&& -0.02\, \r1 + 0.03\, \r2 
  -0.05\, \t1 + 0.01\, \t2 \nn\\*
&& -0.07\, \t3 -0.03\, \t4 -0.051\, \epsilon
  -0.016\, \epsBLM + 0.016\, \epsilon \Lambda \bigg]\,,
\eeqa

We tabulate the shape variables defined in Eq.~(\ref{Rdef}) in
Tables~\ref{tab:1SR0}, \ref{tab:1SR1}, and \ref{tab:1SR2}, and those defined in
Eq.~(\ref{Sdef}) in Tables~\ref{tab:1SS1} and \ref{tab:1SS2} in the $1S$ mass
scheme.  For $S_1$ and $S_2$ we do not show the $E_0$-dependence of the order
$\epsilon\Lambda$ terms, as they are not known.   For all quantities the
coefficients of the expansions are defined as in Eq.~(\ref{expdef}).

\begingroup\squeezetable
\begin{table*}[t]
\caption{Coefficients for $R_0(0,E_0)$ in the $1S$ scheme as a function of
$E_0$.\label{tab:1SR0}}
\begin{tabular}{c||ccccccccccccccccc}
$E_0$  &  $R_0^{(1)}$  &  $R_0^{(2)}$  &  $R_0^{(3)}$  &  $R_0^{(4)}$  &  
  $R_0^{(5)}$  &  $R_0^{(6)}$  &  $R_0^{(7)}$  &  $R_0^{(8)}$  &  
  $R_0^{(9)}$  &  $R_0^{(10)}$  &  $R_0^{(11)}$  &  $R_0^{(12)}$  &  
  $R_0^{(13)}$  &  $R_0^{(14)}$  &  $R_0^{(15)}$  &  $R_0^{(16)}$  &  
  $R_0^{(17)}$  \\ \hline\hline
$0.5$  &  $0.972$  &  $-0.003$  &  $-0.002$  &  $0.$  &  $0.$  &  $-0.01$  & 
    $0.$  &  $0.$  &  $0.$  &  $0.$  &  $0.$  &  $0.$  &  $ 0$  &  $0.$  & 
    $0.$  &  $0.$  &  $0.$ \\ 
$0.7$  &  $0.927$  &  $-0.008$  &  $-0.005$  &  $0.$  &  $-0.01$  &  $ -0.03$ 
    &  $-0.01$  &  $-0.01$  &  $0.$  &  $0.$  &  $-0.01$  &  $0.$  &  $-0.01$ 
    &  $-0.01$  &  $0.001$  &  $0.001$  &  $0.$ \\ 
$0.9$  &  $0.853$  &  $-0.016$  &  $-0.01$  &  $-0.01$  &  $-0.02$  &  $-0.06$ 
    &  $-0.02$  &  $ -0.03$  &  $0.$  &  $0.01$  &  $-0.01$  &  $0.$  & 
    $-0.02$  &  $-0.01$  &  $0.002$  &  $0.001$  &  $0.$ \\ 
$1.1$  &  $0.749$  &  $-0.028$  &  $-0.015$  &  $-0.01$  &  $-0.04$  &  $-0.1$ 
    &  $-0.03$  &  $-0.05$  &  $-0.01$  &  $ 0.01$  &  $-0.02$  &  $0.$  & 
    $-0.03$  &  $-0.02$  &  $0.002$  &  $0.001$  &  $0.$ \\ 
$1.3$  &  $0.615$  &  $-0.043$  &  $-0.022$  &  $-0.01$  &  $-0.06$  &  $
    -0.15$  &  $-0.05$  &  $-0.08$  &  $-0.01$  &  $0.02$  &  $-0.03$  &  $0.$ 
    &  $-0.04$  &  $-0.03$  &  $0.003$  &  $0.002$  &  $ 0.$ \\ 
$1.5$  &  $0.455$  &  $-0.062$  &  $-0.029$  &  $-0.01$  &  $-0.08$  &  $-0.2$ 
    &  $ -0.07$  &  $-0.11$  &  $-0.01$  &  $0.03$  &  $-0.04$  &  $0.$  & 
    $-0.05$  &  $-0.04$  &  $0.003$  &  $0.002$  &  $0.$ \\ 
$1.7$  &  $0.279$  &  $-0.084$  &  $-0.037$  &  $-0.02$  &  $-0.1$  &  $-0.25$ 
    &  $-0.08$  &  $ -0.15$  &  $-0.01$  &  $0.03$  &  $-0.04$  &  $-0.01$  & 
    $-0.06$  &  $-0.05$  &  $0.002$  &  $0.003$  &  $-0.001$ \\

\end{tabular}
\end{table*}
\endgroup

\begingroup\squeezetable
\begin{table*}[t]
\caption{Coefficients for $R_1(E_0)$ in the $1S$ scheme as a function of
$E_0$.\label{tab:1SR1}}
\begin{tabular}{c||ccccccccccccccccc}
$E_0$  &  $R_1^{(1)}$  &  $R_1^{(2)}$  &  $R_1^{(3)}$  &  $R_1^{(4)}$  &  
  $R_1^{(5)}$  &  $R_1^{(6)}$  &  $R_1^{(7)}$  &  $R_1^{(8)}$  &  
  $R_1^{(9)}$  &  $R_1^{(10)}$  &  $R_1^{(11)}$  &  $R_1^{(12)}$  &  
  $R_1^{(13)}$  &  $R_1^{(14)}$  &  $R_1^{(15)}$  &  $R_1^{(16)}$  &  
  $R_1^{(17)}$  \\ \hline\hline
$0$  &  $1.392$  &  $-0.077$  &  $-0.026$  &  $-0.01$  &  $-0.11$  &  $-0.22$ 
    &  $-0.07$  &  $-0.08$  &  $ -0.04$  &  $0.01$  &  $-0.04$  &  $-0.02$  & 
    $-0.05$  &  $-0.05$  &  $0.003$  &  $0.003$  &  $0.$ \\ 
$0.5$  &  $1.422$  &  $-0.076$  &  $-0.025$  &  $-0.01$  &  $-0.11$  &
    $-0.22$  &  $-0.06$  &  $-0.08$  &  $-0.04$  &  $0.01$  &  $-0.04$  & 
    $-0.02$  &  $-0.05$  &  $-0.05$  &  $0.003$  &  $ 0.003$  &  $0.$ \\ 
$0.7$  &  $1.461$  &  $-0.075$  &  $-0.023$  &  $-0.01$  &  $-0.11$  &  
    $-0.21$  &  $-0.06$  &  $-0.08$  &  $-0.04$  &  $0.01$  &  $-0.04$  & 
    $-0.02$  &  $-0.05$  &  $-0.04$  &  $0.002$  &  $ 0.003$  &  $0.$ \\ 
$0.9$  &  $1.517$  &  $-0.074$  &  $ -0.022$  &  $-0.01$  &  $-0.11$  & 
    $-0.2$  &  $-0.06$  &  $-0.08$  &  $-0.04$  &  $0.01$  &  $ -0.04$  & 
    $-0.02$  &  $-0.05$  &  $-0.04$  &  $0.002$  &  $0.003$  &  $0.$ \\ 
$1.1$  &  $1.588$  &  $-0.074$  &  $-0.021$  &  $-0.01$  &  $-0.11$  & 
    $-0.19$  &  $-0.06$  &  $ -0.08$  &  $-0.04$  &  $0.$  &  $-0.04$  & 
    $-0.03$  &  $-0.04$  &  $-0.04$  &  $0.001$  &  $0.003$  &  $0.$ \\ 
$1.3$  &  $1.672$  &  $-0.075$  &  $-0.02$  &  $-0.01$  &  $-0.11$  &  $-0.19$ 
    &  $-0.06$  &  $ -0.07$  &  $-0.05$  &  $0.$  &  $-0.04$  &  $-0.03$  & 
    $-0.04$  &  $-0.04$  &  $0.001$  &  $0.003$  &  $0.$ \\ 
$1.5$  &  $1.767$  &  $-0.077$  &  $-0.02$  &  $-0.01$  &  $-0.12$  &  
    $-0.17$  &  $-0.07$  &  $-0.07$  &  $-0.06$  &  $-0.02$  &  $-0.04$  & 
    $-0.04$  &  $-0.04$  &  $-0.04$  &  $0.001$  &  $ 0.003$  &  $0.$ \\ 
$1.7$  &  $1.872$  &  $-0.08$  &  $-0.021$  &  $-0.01$  &  $-0.14$  &  $-0.16$ 
    &  $-0.1$  &  $ -0.06$  &  $-0.1$  &  $-0.04$  &  $-0.04$  &  $-0.06$  & 
    $-0.03$  &  $-0.03$  &  $0.001$  &  $0.003$  &  $0.$ \\
\end{tabular}
\end{table*}
\endgroup

\begingroup\squeezetable
\begin{table*}[t]
\caption{Coefficients for $R_2(E_0)$ in the $1S$ scheme as a function of $E_0$.
\label{tab:1SR2}}
\begin{tabular}{c||ccccccccccccccccc}
$E_0$  &  $R_2^{(1)}$  &  $R_2^{(2)}$  &  $R_2^{(3)}$  &  $R_2^{(4)}$  &  
  $R_2^{(5)}$  &  $R_2^{(6)}$  &  $R_2^{(7)}$  &  $R_2^{(8)}$  &  
  $R_2^{(9)}$  &  $R_2^{(10)}$  &  $R_2^{(11)}$  &  $R_2^{(12)}$  &  
  $R_2^{(13)}$  &  $R_2^{(14)}$  &  $R_2^{(15)}$  &  $R_2^{(16)}$  &  
  $R_2^{(17)}$  \\ \hline\hline
$0$  &  $2.118$  &  $-0.247$  &  $-0.07$  &  $-0.02$  &  $ -0.36$  &  $-0.68$ 
    &  $-0.19$  &  $-0.21$  &  $-0.15$  &  $0.02$  &  $-0.14$  &  $-0.08$  &  
    $-0.16$  &  $-0.14$  &  $0.008$  &  $0.01$  &  $-0.001$ \\ 
$0.5$  &  $2.175$  &  $-0.247$  &  $-0.069$  &  $-0.02$  &  $ -0.36$  & 
    $-0.68$  &  $-0.19$  &  $-0.22$  &  $-0.15$  &  $0.02$  &  $-0.13$  & 
    $-0.08$  &  $ -0.16$  &  $-0.14$  &  $0.007$  &  $0.009$  &  $-0.001$ \\ 
$0.7$  &  $2.263$  &  $-0.248$  &  $-0.067$  &  $-0.02$  &  $ -0.36$  & 
    $-0.68$  &  $-0.19$  &  $-0.22$  &  $-0.16$  &  $0.01$  &  $-0.13$  & 
    $-0.09$  &  $ -0.15$  &  $-0.14$  &  $0.007$  &  $0.009$  &  $-0.001$ \\ 
$0.9$  &  $2.401$  &  $-0.252$  &  $-0.065$  &  $-0.02$  &  $-0.37$  & 
    $-0.67$  &  $-0.19$  &  $ -0.22$  &  $-0.17$  &  $0.01$  &  $-0.13$  & 
    $-0.1$  &  $-0.15$  &  $-0.14$  &  $0.005$  &  $0.009$  &  $-0.001$ \\ 
$1.1$  &  $2.593$  &  $-0.259$  &  $-0.064$  &  $-0.02$  &  $ -0.38$  & 
    $-0.67$  &  $-0.19$  &  $-0.23$  &  $-0.18$  &  $-0.01$  &  $ -0.13$  & 
    $-0.11$  &  $-0.14$  &  $-0.14$  &  $0.004$  &  $0.009$  &  $-0.001$ \\ 
$1.3$  &  $2.842$  &  $-0.271$  &  $-0.063$  &  $-0.02$  &  $ -0.41$  & 
    $-0.66$  &  $-0.21$  &  $-0.23$  &  $-0.21$  &  $-0.03$  &  $-0.14$  & 
    $-0.13$  &  $ -0.14$  &  $-0.14$  &  $0.003$  &  $0.009$  &  $-0.001$ \\ 
$1.5$  &  $3.15$  &  $-0.288$  &  $-0.066$  &  $-0.02$  &  $-0.46$  &  
    $-0.64$  &  $-0.24$  &  $-0.23$  &  $-0.28$  &  $-0.07$  &  $-0.14$  &  
    $-0.18$  &  $-0.13$  &  $-0.14$  &  $0.003$  &  $0.011$  &  $-0.001$ \\ 
$1.7$  &  $3.518$  &  $-0.311$  &  $-0.072$  &  $-0.02$  &  $ -0.58$  & 
    $-0.62$  &  $-0.35$  &  $-0.21$  &  $-0.43$  &  $-0.16$  &  $ -0.16$  & 
    $-0.26$  &  $-0.12$  &  $-0.13$  &  $0.004$  &  $0.013$  &  $0.$ \\
\end{tabular}
\end{table*}
\endgroup

\begingroup\squeezetable
\begin{table*}[t]
\caption{Coefficients for $S_1(E_0)$ in the $1S$ scheme as a function of $E_0$.
\label{tab:1SS1}}
\begin{tabular}{c||ccccccccccccccccc}
$E_0$  &  $S_1^{(1)}$  &  $S_1^{(2)}$  &  $S_1^{(3)}$  &  $S_1^{(4)}$  &  
  $S_1^{(5)}$  &  $S_1^{(6)}$  &  $S_1^{(7)}$  &  $S_1^{(8)}$  &  
  $S_1^{(9)}$  &  $S_1^{(10)}$  &  $S_1^{(11)}$  &  $S_1^{(12)}$  &  
  $S_1^{(13)}$  &  $S_1^{(14)}$  &  $S_1^{(15)}$  &  $S_1^{(16)}$  &  
  $S_1^{(17)}$  \\ \hline\hline
$0$  &  $0.832$  &  $1.633$  &  $0.416$  &  $0.13$  &  $1.49$  &  $-0.36$  & 
   $0.75$  &  $0.$  &  $0.46$  &  $-0.24$  &  $0.53$  &  $ 0.25$  &  $0.5$  & 
   $0.14$  &  $0.044$  &  $-0.025$  &  $0.025$ \\ 
$0.5$  &  $0.82$  &  $1.609$  &  $0.409$  &  $0.12$  &  $1.5$  &  $-0.32$  & 
   $0.75$  &  $0.02$  &  $0.48$  &  $-0.24$  &  $0.54$  &  $ 0.26$  &  $0.5$ 
   &  $0.14$  &  $0.039$  &  $-0.028$  &  --- \\ 
$0.7$  &  $0.805$  &  $1.578$  &  $0.398$  &  $0.12$  &  $1.52$  &  $-0.26$  & 
   $0.77$  &  $0.05$  &  $0.5$  &  $-0.23$  &  $0.54$  &  $ 0.27$  &  $0.5$  & 
   $0.16$  &  $0.032$  &  $-0.031$  &  --- \\ 
$0.9$  &  $0.784$  &  $1.533$  &  $0.38$  &  $0.11$  &  $1.56$  &  $-0.16$  & 
   $0.79$  &  $0.12$  &  $0.55$  &  $-0.22$  &  $0.55$  &  $ 0.3$  &  $0.51$ 
   &  $0.18$  &  $0.023$  &  $-0.035$  &  --- \\ 
$1.1$  &  $0.759$  &  $1.479$  &  $0.354$  &  $0.1$  &  $1.63$  &  $-0.02$  & 
   $0.83$  &  $0.22$  &  $0.63$  &  $-0.2$  &  $0.57$  &  $ 0.34$  &  $0.52$ 
   &  $0.2$  &  $0.011$  &  $-0.04$  &  --- \\ 
$1.3$  &  $0.734$  &  $1.42$  &  $0.319$  &  $0.09$  &  $1.74$  &  $0.18$  & 
   $0.91$  &  $0.38$  &  $0.77$  &  $-0.16$  &  $0.59$  &  $ 0.41$  &  $0.54$ 
   &  $0.24$  &  $-0.002$  &  $-0.046$  &  --- \\ 
$1.5$  &  $0.716$  &  $1.371$  &  $0.277$  &  $0.06$  &  $1.97$  &  $0.45$  & 
   $1.07$  &  $0.65$  &  $1.03$  &  $-0.06$  &  $0.64$  &  $ 0.55$  &  $0.56$ 
   &  $0.3$  &  $-0.018$  &  $-0.054$  &  --- \\ 
$1.7$  &  $0.72$  &  $1.368$  &  $0.254$  &  $0.05$  &  $2.49$  &  $0.84$  & 
   $1.59$  &  $1.13$  &  $1.64$  &  $0.22$  &  $0.76$  &  $ 0.86$  &  $0.6$  & 
   $0.38$  &  $-0.035$  &  $-0.066$  &  --- \\
\end{tabular}
\end{table*}
\endgroup

\begingroup\squeezetable
\begin{table*}[t]
\caption{Coefficients for $S_2(E_0)$ in the $1S$ scheme as a function of $E_0$.
\label{tab:1SS2}}
\begin{tabular}{c||ccccccccccccccccc}
$E_0$  &  $S_2^{(1)}$  &  $S_2^{(2)}$  &  $S_2^{(3)}$  &  $S_2^{(4)}$  &  
  $S_2^{(5)}$  &  $S_2^{(6)}$  &  $S_2^{(7)}$  &  $S_2^{(8)}$  &  
  $S_2^{(9)}$  &  $S_2^{(10)}$  &  $S_2^{(11)}$  &  $S_2^{(12)}$  &  
  $S_2^{(13)}$  &  $S_2^{(14)}$  &  $S_2^{(15)}$  &  $S_2^{(16)}$  &  
  $S_2^{(17)}$  \\ \hline\hline
$0$  &  $0.125$  &  $0.472$  &  $0.531$  &  $0.16$  &  $-4.43$  &  $-0.68$  & 
   $-1.04$  &  $-1.6$  &  $-5.46$  &  $1.07$  &  $ -0.94$  &  $-2.8$  & 
   $-0.05$  &  $-0.13$  &  $0.381$  &  $-0.428$  &  $0.171$ \\ 
$0.5$  &  $0.123$  &  $0.467$  &  $0.524$  &  $0.16$  &  $-4.34$  &  $-0.66$ 
   &  $-0.99$  &  $-1.55$  &  $-5.53$  &  $0.96$  &  $ -0.93$  &  $-2.74$  & 
   $-0.05$  &  $-0.12$  &  $0.405$  &  $-0.42$  &  --- \\ 
$0.7$  &  $0.123$  &  $0.465$  &  $0.521$  &  $0.16$  &  $-4.23$  &  $-0.64$ 
   &  $-0.91$  &  $-1.5$  &  $-5.64$  &  $0.81$  &  $ -0.9$  &  $-2.67$  & 
   $-0.05$  &  $-0.12$  &  $0.448$  &  $-0.408$  &  --- \\ 
$0.9$  &  $0.124$  &  $0.468$  &  $0.524$  &  $0.16$  &  $-4.08$  &  $-0.62$ 
   &  $-0.78$  &  $-1.43$  &  $-5.85$  &  $0.59$  &  $ -0.87$  &  $-2.58$  & 
   $-0.05$  &  $-0.11$  &  $0.526$  &  $-0.391$  &  --- \\ 
$1.1$  &  $0.126$  &  $0.477$  &  $0.533$  &  $0.16$  &  $-3.89$  &  $-0.6$  & 
   $-0.6$  &  $-1.36$  &  $-6.2$  &  $0.28$  &  $ -0.83$  &  $-2.46$  & 
   $-0.05$  &  $-0.11$  &  $0.661$  &  $-0.37$  &  --- \\ 
$1.3$  &  $0.128$  &  $0.486$  &  $0.546$  &  $0.17$  &  $-3.69$  &  $-0.57$ 
   &  $-0.35$  &  $-1.28$  &  $-6.79$  &  $-0.11$  &  $ -0.79$  &  $-2.33$  & 
   $-0.05$  &  $-0.1$  &  $0.892$  &  $-0.344$  &  --- \\ 
$1.5$  &  $0.128$  &  $0.487$  &  $0.55$  &  $0.18$  &  $-3.5$  &  $-0.53$  & 
   $-0.04$  &  $-1.19$  &  $-7.88$  &  $-0.61$  &  $ -0.75$  &  $-2.21$  & 
   $-0.05$  &  $-0.1$  &  $1.328$  &  $-0.311$  &  --- \\ 
$1.7$  &  $0.12$  &  $0.454$  &  $0.509$  &  $0.16$  &  $-3.46$  &  $-0.49$  & 
   $0.16$  &  $-1.08$  &  $-10.34$  &  $-1.34$  &  $ -0.74$  &  $-2.18$  & 
   $-0.05$  &  $-0.09$  &  $2.345$  &  $-0.273$  &  --- \\
\end{tabular}
\end{table*}
\endgroup

For the $B\to X_s\gamma$ shape variables defined in Eq.~(\ref{Tdef}), only
$T_i^{(15)}$, $T_i^{(16)}$, and $T_i^{(17)}$ are functions of $E_0$, once
$m_B/2-E_0 \gg \lqcd$.  For the other $T$'s in the $1S$ scheme we find
\beqa\label{upsbsg1}
&& T_1^{(1)} = \frac{m_\Upsilon}4, \qquad
  T_1^{(2)} = -\frac12, \qquad
  T_1^{(3)} = T_1^{(4)} = 0, \qquad
  T_1^{(5)} = -0.05, \qquad
  T_1^{(6)} = -0.16, \nn\\*
&& T_1^{(7)} = -0.01, \qquad
  T_1^{(8)} = -0.03, \qquad
  T_1^{(9)} = -0.02, \qquad
  T_1^{(10)} = 0.18, \nn\\*
&& T_1^{(11)} = T_1^{(13)} = -0.01, \qquad
  T_1^{(12)} = T_1^{(14)} = -0.03,
\eeqa
and
\beqa\label{upsbsg2}
&& T_2^{(1)} = T_2^{(2)} = T_2^{(3)} = T_2^{(4)} = 
  T_2^{(6)} = T_2^{(7)} = T_2^{(8)} = T_2^{(13)} = T_2^{(14)} = 0, \\*
&&T_2^{(5)} = -\frac1{12}, \qquad
  T_2^{(9)} = -0.04, \qquad
  T_2^{(10)} = -T_2^{(12)} = 0.05, \qquad
  T_2^{(11)} = -0.02. \nn
\eeqa
The remaining, $E_0$-dependent coefficients of the perturbative corrections are
listed in Table~\ref{tab:upsbsg}.

\begingroup\squeezetable
\begin{table*}[t]
\caption{Perturbative coefficients for $T_1(E_0)$ and $T_2(E_0)$ in the $1S$
scheme as a function of $E_0$.\label{tab:upsbsg}}
\begin{tabular}{c||ccc||ccc}
$E_0$  &  $T_1^{(15)}$  &  $T_1^{(16)}$  &  $T_1^{(17)}$  
  &  $T_2^{(15)}$  &  $T_2^{(16)}$  &  $T_2^{(17)}$  \\ \hline\hline
$1.7$  &  $-0.043$  &  $-0.017$  &  $0.016$
  &  $0.016$  &  $0.011$  &  $-0.014$ \\ 
$1.8$  &  $-0.038$  &  $-0.014$  &  $0.021$
  &  $0.012$  &  $0.009$  &  $-0.014$ \\ 
$1.9$  &  $-0.032$  &  $-0.011$  &  $0.026$  
  &  $0.01$  &  $0.007$  &  $-0.014$ \\ 
$2$  &  $-0.025$  &  $-0.006$  &  $0.033$  
  &  $0.007$  &  $0.006$  &  $-0.013$ \\ 
$2.1$  &  $-0.017$  &  $-0.001$  &  $0.042$  
  &  $0.004$  &  $0.004$  &  $-0.012$ \\ 
$2.2$  &  $-0.007$  &  $0.008$  &  $0.056$  
  &  $0.002$  &  $0.002$  &  $-0.01$ \\ 
\end{tabular}
\end{table*}
\endgroup

\subsection{The PS mass scheme}

The expressions for the $B\to X_c\ell\bar\nu$ decay rate and the shape
variables in the PS scheme are almost identical to Eq.~(\ref{upswidth}),
Tables~\ref{tab:1SR0}--\ref{tab:1SS2}, and Eqs.~(\ref{upsbsg1}) and
(\ref{upsbsg2}), because we choose to expand \mbps\ about $m_\Upsilon/2$ as
well.  The difference in the $B\to X_c\ell\bar\nu$ rate compared with 
Eq.~(\ref{upswidth}) is that the perturbation series is replaced by $-0.020\,
\epsilon -0.003\, \epsBLM + 0.025\, \epsilon \Lambda$, and of course, the
meaning of $\Lambda$ changes from $\lups$ to $\lps$.  

Next we tabulate the coefficients of the perturbation series of the shape
variables defined in Eqs.~(\ref{Rdef}) and (\ref{Sdef}), that differ from the
entries in Tables~\ref{tab:1SR0}--\ref{tab:1SS2}, in Table~\ref{tab:PSlepton}
in the PS mass scheme.  For $S_1$ and $S_2$ we do not show in the tables the
order $\epsilon\Lambda$ terms again as their $E_0$-dependence is not known. 
For all quantities the coefficients of the expansions are defined as in
Eq.~(\ref{expdef}).

\begingroup\squeezetable
\begin{table*}[t]
\caption{Perturbative coefficients for $R_0(0,E_0)$, $R_1(E_0)$, $R_2(E_0)$,
$S_1(E_0)$, and $S_2(E_0)$ in the PS scheme, that differ from the results in
the $1S$ scheme, as a function of $E_0$.\label{tab:PSlepton}}
\begin{tabular}{c||ccc||ccc||ccc||ccc||ccc}
$E_0$  &  $R_0^{(15)}$  &  $R_0^{(16)}$  &  $R_0^{(17)}$  
  &  $R_1^{(15)}$  &  $R_1^{(16)}$  &  $R_1^{(17)}$  
  &  $R_2^{(15)}$  &  $R_2^{(16)}$  &  $R_2^{(17)}$ 
  &  $S_1^{(15)}$  &  $S_1^{(16)}$  &  $S_1^{(17)}$  
  &  $S_2^{(15)}$  &  $S_2^{(16)}$  &  $S_2^{(17)}$   
  \\ \hline\hline
$0$    &  ---  &  ---  &  ---  
  &  $0.013$  &   $0.007$  &  $0.008$ 
  &  $0.041$  &  $0.024$  &  $0.021$ 
  &  $-0.178$  &  $-0.106$  &  $-0.106$ 
  &  $0.317$  &  $-0.452$  &  $0.022$ \\
$0.5$  &  $0.001$  &  $0.$  &  $0.001$ 
  &  $0.013$  &  $0.007$  &  $0.007$ 
  &  $0.041$  &  $0.023$  &  $0.02$ 
  &  $-0.18$  &  $-0.108$  &  ---
  &  $0.342$  &  $-0.443$  &  --- \\ 
$0.7$  &  $0.002$  &  $0.001$  &  $0.002$ 
  &  $0.012$  &  $0.007$  &  $0.007$ 
  &  $0.04$  &  $0.023$  &  $0.02$ 
  &  $-0.182$  &  $-0.109$  &  ---
  &  $0.385$  &  $-0.432$  &  --- \\ 
$0.9$  &  $0.004$  &  $0.002$  &  $0.003$ 
  &  $0.012$  &  $0.007$  &  $0.007$ 
  &  $0.04$  &  $0.023$  &  $0.019$ 
  &  $-0.186$  &  $-0.111$  &  --- 
  &  $0.462$  &  $-0.415$  &  --- \\ 
$1.1$  &  $0.006$  &  $0.003$  &  $0.005$ 
  &  $0.011$  &  $0.007$  &  $0.006$ 
  &  $0.039$  &  $0.024$  &  $0.019$ 
  &  $-0.19$  &  $-0.114$  &  --- 
  &  $0.596$  &  $-0.393$  &  --- \\ 
$1.3$  &  $0.009$  &  $0.004$  &  $0.007$ 
  &  $0.011$  &  $0.007$  &  $0.006$ 
  &  $0.04$  &  $0.025$  &  $0.019$ 
  &  $-0.195$  &  $-0.116$  &  ---
  &  $0.826$  &  $-0.368$  &  --- \\ 
$1.5$  &  $0.011$  &  $0.006$  &  $0.009$ 
  &  $0.011$  &  $0.007$  &  $0.006$ 
  &  $0.042$  &  $0.027$  &  $0.02$ 
  &  $-0.205$  &  $-0.122$  &  ---
  &  $1.262$  &  $-0.335$  &  --- \\ 
$1.7$  &  $0.013$  &  $0.007$  &  $0.01$ 
  &  $0.012$  &  $0.008$  &  $0.007$ 
  &  $0.046$  &  $0.031$  &  $0.023$ 
  &  $-0.221$  &  $-0.135$  &  ---
  &  $2.283$  &  $-0.296$  &  --- \\
\end{tabular}
\end{table*}
\endgroup

For the $B\to X_s\gamma$ shape variables defined in Eq.~(\ref{Tdef}),
the expressions for $T_2$ are identical in the $1S$ and PS schemes, and so only
$T_1^{(15)}$, $T_1^{(16)}$, and $T_1^{(17)}$ differ between these two schemes. 
The results for these coefficients in the PS scheme are shown in
Table~\ref{tab:psbsg}.

\begingroup\squeezetable
\begin{table*}[t]
\caption{Perturbative coefficients for $T_1(E_0)$ in the PS
scheme as a function of $E_0$.\label{tab:psbsg}}
\begin{tabular}{c||ccc}
$E_0$  &  $T_1^{(15)}$  &  $T_1^{(16)}$  &  $T_1^{(17)}$  \\ \hline\hline
$1.7$  &  $0.025$  &  $0.011$  &  $0.022$ \\ 
$1.8$  &  $0.03$  &  $0.014$  &  $0.026$ \\ 
$1.9$  &  $0.036$  &  $0.018$  &  $0.032$ \\ 
$2$  &  $0.043$  &  $0.022$  &  $0.038$ \\ 
$2.1$  &  $0.051$  &  $0.028$  &  $0.047$ \\ 
$2.2$  &  $0.061$  &  $0.036$  &  $0.062$ \\ 
\end{tabular}
\end{table*}
\endgroup

\subsection{\boldmath The \MSbar\ mass scheme}

The $B\to X_c\ell\bar\nu$ decay width in the \MSbar\ scheme is given by
\beqa\label{mswidth}
\Gamma(B\to X_c\ell\bar\nu) &=& {G_F^2\, |V_{cb}|^2\over 192\pi^3}\,
  (4.2\,\GeV)^5\, \bigg[ 0.733
  -0.464\, \Lambda -0.036\, \Lambda^2 + 0.01\, \Lambda^3 \nn\\*
&& -0.22\, \l1 -0.22\, \l2 
  -0.04\, \l1 \Lambda + 0.1\, \l2 \Lambda \nn\\*
&& -0.01\, \r1 + 0.05\, \r2 
  -0.16\, \t1 + 0.01\, \t2 \nn\\*
&& -0.18\, \t3 -0.05\, \t4 +0.085\, \epsilon
  +0.065\, \epsBLM + 0.022\, \epsilon \Lambda \bigg]\,,
\eeqa

We tabulate the shape variables defined in Eq.~(\ref{Rdef}) in
Tables~\ref{tab:MSR0}, \ref{tab:MSR1}, and \ref{tab:MSR2}, and those defined in
Eq.~(\ref{Sdef}) in Tables~\ref{tab:MSS1} and \ref{tab:MSS2} in the \MSbar\
mass scheme.  For $S_1$ and $S_2$ we do not show the $E_0$-dependence of the
order $\epsilon\Lambda$ terms, as they are not known.   For all quantities the
coefficients of the expansions are defined as in Eq.~(\ref{expdef}).

\begingroup\squeezetable
\begin{table*}[t]
\caption{Coefficients for $R_0(0,E_0)$ in the \MSbar\ scheme as a function of
$E_0$.\label{tab:MSR0}}
\begin{tabular}{c||ccccccccccccccccc}
$E_0$  &  $R_0^{(1)}$  &  $R_0^{(2)}$  &  $R_0^{(3)}$  &  $R_0^{(4)}$  &  
  $R_0^{(5)}$  &  $R_0^{(6)}$  &  $R_0^{(7)}$  &  $R_0^{(8)}$  &  
  $R_0^{(9)}$  &  $R_0^{(10)}$  &  $R_0^{(11)}$  &  $R_0^{(12)}$  &  
  $R_0^{(13)}$  &  $R_0^{(14)}$  &  $R_0^{(15)}$  &  $R_0^{(16)}$  &  
  $R_0^{(17)}$  \\ \hline\hline
$0.5$  &  $0.969$  &  $-0.007$  &  $-0.005$  &  $0.$  &  $-0.01$  &  $-0.01$ 
    &  $-0.01$  &  $-0.01$  &  $0.$  &  $0.$  &  $-0.01$  &  $0.$  &  $-0.01$ 
    &  $0.$  &  $0.003$  &  $0.001$  &  $0.003$ \\ 
$0.7$  &  $0.92$  &  $-0.017$  &  $-0.013$  &  $-0.01$  &  $-0.02$  &  $-0.04$ 
    &  $-0.02$  &  $ -0.02$  &  $0.$  &  $0.01$  &  $-0.01$  &  $0.$  & 
    $-0.02$  &  $-0.01$  &  $0.007$  &  $0.004$  &  $0.008$ \\ 
$0.9$  &  $0.841$  &  $-0.033$  &  $-0.025$  &  $-0.02$  &  $-0.04$  & 
    $-0.07$  &  $-0.04$  &  $-0.05$  &  $0.$  &  $0.01$  &  $ -0.03$  &  $0.$ 
    &  $-0.03$  &  $-0.02$  &  $0.013$  &  $0.007$  &  $0.015$ \\ 
$1.1$  &  $0.729$  &  $-0.054$  &  $-0.04$  &  $-0.03$  &  $-0.06$  &  $-0.13$ 
    &  $-0.07$  &  $-0.09$  &  $0.$  &  $0.02$  &  $ -0.05$  &  $0.$  & 
    $-0.06$  &  $-0.03$  &  $0.021$  &  $0.012$  &  $0.023$\\
$1.3$  &  $0.584$  &  $-0.08$  &  $-0.056$  &  $-0.04$  &  $-0.1$  &  $-0.2$ 
    &  $-0.11$  &  $-0.14$  &  $0.$  &  $0.03$  &  $-0.07$  &  $0.$  & 
    $-0.08$  &  $-0.05$  &  $0.031$  &  $0.017$  &  $0.032$ \\ 
$1.5$  &  $0.411$  &  $-0.11$  &  $-0.071$  &  $-0.05$  &  $-0.13$  &  
    $-0.29$  &  $-0.15$  &  $-0.22$  &  $0.$  &  $0.04$  &  $-0.09$  &  $0.$  & 
    $-0.11$  &  $-0.07$  &  $0.041$  &  $ 0.024$  &  $0.039$ \\ 
$1.7$  &  $0.221$  &  $-0.145$  &  $-0.086$  &  $-0.05$  &  $-0.16$  &  
    $-0.36$  &  $-0.18$  &  $-0.3$  &  $0.01$  &  $0.04$  &  $-0.11$  & 
    $-0.01$  &  $ -0.13$  &  $-0.09$  &  $0.052$  &  $0.032$  &  $0.046$ \\
\end{tabular}
\end{table*}
\endgroup

\begingroup\squeezetable
\begin{table*}[t]
\caption{Coefficients for $R_1(E_0)$ in the \MSbar\ scheme as a function of
$E_0$.\label{tab:MSR1}}
\begin{tabular}{c||ccccccccccccccccc}
$E_0$  &  $R_1^{(1)}$  &  $R_1^{(2)}$  &  $R_1^{(3)}$  &  $R_1^{(4)}$  &  
  $R_1^{(5)}$  &  $R_1^{(6)}$  &  $R_1^{(7)}$  &  $R_1^{(8)}$  &  
  $R_1^{(9)}$  &  $R_1^{(10)}$  &  $R_1^{(11)}$  &  $R_1^{(12)}$  &  
  $R_1^{(13)}$  &  $R_1^{(14)}$  &  $R_1^{(15)}$  &  $R_1^{(16)}$  &  
  $R_1^{(17)}$  \\ \hline\hline
$0$  &  $1.342$  &  $-0.117$  &  $-0.054$  &  $-0.03$  &  $-0.16$  &  $-0.27$ 
    &  $-0.12$  &  $ -0.12$  &  $-0.03$  &  $0.01$  &  $-0.09$  &  $-0.03$  & 
    $-0.1$  &  $-0.07$  &  $ 0.043$  &  $0.026$  &  $0.027$ \\ 
$0.5$  &  $1.373$  &  $-0.113$  &  $-0.05$  &  $-0.03$  &  $-0.15$  &  $-0.27$ 
    &  $ -0.12$  &  $-0.12$  &  $-0.04$  &  $0.01$  &  $-0.09$  &  $-0.03$  & 
    $ -0.1$  &  $-0.06$  &  $0.042$  &  $0.025$  &  $0.025$ \\ 
$0.7$  &  $1.413$  &  $-0.11$  &  $-0.047$  &  $-0.02$  &  $-0.15$  &  $-0.26$ 
    &  $-0.11$  &  $ -0.12$  &  $-0.04$  &  $0.01$  &  $-0.09$  &  $-0.03$  & 
    $-0.1$  &  $-0.06$  &  $0.04$  &  $ 0.024$  &  $0.023$ \\ 
$0.9$  &  $1.47$  &  $-0.106$  &  $-0.043$  &  $-0.02$  &  $-0.15$  &  
    $-0.26$  &  $-0.11$  &  $-0.12$  &  $-0.04$  &  $0.01$  &  $-0.08$  & 
    $-0.03$  &  $-0.09$  &  $ -0.06$  &  $0.039$  &  $0.024$  &  $0.021$ \\ 
$1.1$  &  $1.542$  &  $-0.104$  &  $-0.039$  &  $-0.02$  &  $-0.15$  & 
    $-0.25$  &  $-0.11$  &  $ -0.12$  &  $-0.05$  &  $0.$  &  $-0.08$  & 
    $-0.04$  &  $-0.09$  &  $-0.06$  &  $ 0.037$  &  $0.023$  &  $0.019$ \\ 
$1.3$  &  $1.626$  &  $-0.103$  &  $-0.036$  &  $-0.02$  &  $-0.15$  &  
    $-0.24$  &  $-0.11$  &  $-0.12$  &  $-0.06$  &  $-0.01$  &  $-0.08$  &  
    $-0.05$  &  $-0.08$  &  $-0.06$  &  $0.037$  &  $0.023$  &  $0.017$ \\ 
$1.5$  &  $1.72$  &  $-0.105$  &  $-0.035$  &  $-0.01$  &  $-0.17$  &  $-0.22$ 
    &  $-0.13$  &  $ -0.12$  &  $-0.08$  &  $-0.03$  &  $-0.08$  &  $-0.07$  & 
    $-0.07$  &  $-0.05$  &  $ 0.037$  &  $0.024$  &  $0.016$ \\ 
$1.7$  &  $1.823$  &  $-0.109$  &  $-0.036$  &  $-0.01$  &  $-0.22$  &  $-0.2$ 
    &  $-0.22$  &  $ -0.1$  &  $-0.16$  &  $-0.08$  &  $-0.08$  &  $-0.11$  & 
    $-0.06$  &  $-0.05$  &  $0.039$  &  $0.025$  &  $0.017$ \\
\end{tabular}
\end{table*}
\endgroup

\begingroup\squeezetable
\begin{table*}[t]
\caption{Coefficients for $R_2(E_0)$ in the \MSbar\ scheme as a function of
$E_0$.\label{tab:MSR2}}
\begin{tabular}{c||ccccccccccccccccc}
$E_0$  &  $R_2^{(1)}$  &  $R_2^{(2)}$  &  $R_2^{(3)}$  &  $R_2^{(4)}$  &  
  $R_2^{(5)}$  &  $R_2^{(6)}$  &  $R_2^{(7)}$  &  $R_2^{(8)}$  &  
  $R_2^{(9)}$  &  $R_2^{(10)}$  &  $R_2^{(11)}$  &  $R_2^{(12)}$  &  
  $R_2^{(13)}$  &  $R_2^{(14)}$  &  $R_2^{(15)}$  &  $R_2^{(16)}$  &  
  $R_2^{(17)}$  \\ \hline\hline
$0$  &  $1.963$  &  $-0.35$  &  $-0.136$  &  $-0.07$  &  $-0.49$  &  $-0.82$ 
    &  $-0.31$  &  $-0.31$  &  $ -0.14$  &  $0.02$  &  $-0.27$  &  $-0.11$  & 
    $-0.3$  &  $-0.2$  &  $0.129$  &  $0.077$  &  $0.063$ \\
$0.5$  &  $2.02$  &  $-0.348$  &  $-0.132$  &  $-0.06$  &  $-0.49$  &  
    $-0.82$  &  $-0.31$  &  $-0.31$  &  $-0.15$  &  $0.02$  &  $-0.27$  & 
    $-0.11$  &  $-0.3$  &  $-0.2$  &  $0.127$  &  $0.077$  &  $0.061$ \\ 
$0.7$  &  $2.108$  &  $-0.346$  &  $-0.127$  &  $-0.06$  &  $-0.49$  &  
    $-0.82$  &  $-0.32$  &  $-0.32$  &  $-0.15$  &  $0.01$  &  $-0.27$  &  
    $-0.12$  &  $-0.3$  &  $-0.2$  &  $0.126$  &  $0.077$  &  $0.058$ \\ 
$0.9$  &  $2.245$  &  $-0.346$  &  $-0.122$  &  $-0.06$  &  $-0.5$  &  
    $-0.82$  &  $-0.32$  &  $-0.33$  &  $-0.17$  &  $0.$  &  $-0.27$  & 
    $-0.13$  &  $-0.29$  &  $-0.19$  &  $ 0.125$  &  $0.077$  &  $0.054$ \\ 
$1.1$  &  $2.435$  &  $-0.349$  &  $-0.115$  &  $-0.05$  &  $-0.52$  & 
   $-0.82$  &  $ -0.33$  &  $-0.35$  &  $-0.19$  &  $-0.02$  &  $-0.27$  & 
   $-0.15$  &  $-0.28$  &  $-0.19$  &  $0.125$  &  $ 0.078$  &  $0.05$ \\ 
$1.3$  &  $2.678$  &  $-0.359$  &  $-0.11$  &  $-0.05$  &  $-0.55$  &  
    $-0.81$  &  $-0.36$  &  $-0.36$  &  $ -0.24$  &  $-0.05$  &  $-0.27$  & 
    $-0.19$  &  $-0.27$  &  $-0.19$  &  $0.127$  &  $ 0.081$  &  $0.046$ \\ 
$1.5$  &  $2.975$  &  $-0.378$  &  $-0.11$  &  $-0.04$  &  $-0.63$  &  $-0.8$ 
    &  $ -0.43$  &  $-0.36$  &  $-0.34$  &  $-0.12$  &  $ -0.27$  &  $-0.26$ 
    &  $-0.25$  &  $-0.19$  &  $0.134$  &  $0.086$  &  $0.045$ \\ 
$1.7$  &  $3.329$  &  $-0.409$  &  $-0.119$  &  $-0.04$  &  $-0.85$  &  
    $-0.75$  &  $-0.78$  &  $-0.32$  &  $-0.65$  &  $-0.31$  &  $-0.31$  & 
    $-0.44$  &  $-0.22$  &  $ -0.18$  &  $0.147$  &  $0.096$  &  $ 0.053$ \\
\end{tabular}
\end{table*}
\endgroup

\begingroup\squeezetable
\begin{table*}[t]
\caption{Coefficients for $S_1(E_0)$ in the \MSbar\ scheme as a function of
$E_0$.\label{tab:MSS1}}
\begin{tabular}{c||ccccccccccccccccc}
$E_0$  &  $S_1^{(1)}$  &  $S_1^{(2)}$  &  $S_1^{(3)}$  &  $S_1^{(4)}$  &  
  $S_1^{(5)}$  &  $S_1^{(6)}$  &  $S_1^{(7)}$  &  $S_1^{(8)}$  &  
  $S_1^{(9)}$  &  $S_1^{(10)}$  &  $S_1^{(11)}$  &  $S_1^{(12)}$  &  
  $S_1^{(13)}$  &  $S_1^{(14)}$  &  $S_1^{(15)}$  &  $S_1^{(16)}$  &  
  $S_1^{(17)}$  \\ \hline\hline
$0$  &  $1.837$  &  $2.216$  &  $0.729$  &  $0.3$  &  $2.$  &  $-0.31$  & 
   $1.24$  &  $0.21$  &  $0.43$  &  $-0.26$  &  $1.05$  &  $ 0.39$  &  $1.$  & 
   $0.23$  &  $-0.711$  &  $-0.456$  &  $-0.297$ \\ 
$0.5$  &  $1.811$  &  $2.181$  &  $0.715$  &  $0.3$  &  $2.02$  &  $-0.26$  & 
   $1.26$  &  $0.24$  &  $0.45$  &  $-0.26$  &  $1.06$  &  $ 0.41$  &  $1.$  & 
   $0.24$  &  $-0.707$  &  $-0.452$  &  --- \\ 
$0.7$  &  $1.775$  &  $2.134$  &  $0.695$  &  $0.29$  &  $2.05$  &  $-0.17$  & 
   $1.29$  &  $0.3$  &  $0.49$  &  $-0.25$  &  $1.06$  &  $ 0.43$  &  $1.01$ 
   &  $0.26$  &  $-0.7$  &  $-0.446$  &  --- \\ 
$0.9$  &  $1.724$  &  $2.064$  &  $0.664$  &  $0.28$  &  $2.11$  &  $-0.03$  & 
   $1.34$  &  $0.41$  &  $0.57$  &  $-0.23$  &  $ 1.08$  &  $0.46$  &  $1.02$ 
   &  $0.29$  &  $-0.691$  &  $-0.437$  &  --- \\ 
$1.1$  &  $1.662$  &  $1.971$  &  $0.615$  &  $0.26$  &  $2.21$  &  $0.18$  & 
   $1.43$  &  $0.59$  &  $0.7$  &  $-0.19$  &  $1.11$  &  $ 0.53$  &  $1.05$ 
   &  $0.34$  &  $-0.678$  &  $-0.424$  &  --- \\ 
$1.3$  &  $1.593$  &  $1.858$  &  $0.542$  &  $0.23$  &  $2.38$  &  $0.5$  & 
   $1.6$  &  $0.9$  &  $0.92$  &  $-0.12$  &  $1.17$  &  $ 0.63$  &  $1.09$  & 
   $0.41$  &  $-0.664$  &  $-0.408$  &  --- \\ 
$1.5$  &  $1.532$  &  $1.735$  &  $0.434$  &  $0.16$  &  $2.74$  &  $0.98$  & 
   $2.$  &  $1.46$  &  $1.38$  &  $0.07$  &  $1.28$  &  $ 0.85$  &  $1.17$  & 
   $0.52$  &  $-0.66$  &  $-0.391$  &  --- \\ 
$1.7$  &  $1.524$  &  $1.684$  &  $0.351$  &  $0.08$  &  $3.76$  &  $1.81$  & 
   $3.64$  &  $2.88$  &  $2.61$  &  $0.71$  &  $1.59$  &  $ 1.47$  &  $1.34$ 
   &  $0.72$  &  $-0.698$  &  $-0.396$  &  --- \\
\end{tabular}
\end{table*}
\endgroup

\begingroup\squeezetable
\begin{table*}[t]
\caption{Coefficients for $S_2(E_0)$ in the \MSbar\ scheme as a function of
$E_0$.\label{tab:MSS2}}
\begin{tabular}{c||ccccccccccccccccc}
$E_0$  &  $S_2^{(1)}$  &  $S_2^{(2)}$  &  $S_2^{(3)}$  &  $S_2^{(4)}$  &  
  $S_2^{(5)}$  &  $S_2^{(6)}$  &  $S_2^{(7)}$  &  $S_2^{(8)}$  &  
  $S_2^{(9)}$  &  $S_2^{(10)}$  &  $S_2^{(11)}$  &  $S_2^{(12)}$  &  
  $S_2^{(13)}$  &  $S_2^{(14)}$  &  $S_2^{(15)}$  &  $S_2^{(16)}$  &  
  $S_2^{(17)}$  \\ \hline\hline
$0$  &  $0.549$  &  $1.175$  &  $0.78$  &  $0.09$  &  $-5.13$  &  $-1.86$  & 
   $-1.75$  &  $-3.01$  &  $-6.91$  &  $0.7$  &  $ -1.34$  &  $-3.5$  & 
   $-0.3$  &  $-0.39$  &  $0.085$  &  $-1.169$  &  $-0.187$ \\ 
$0.5$  &  $0.542$  &  $1.16$  &  $0.769$  &  $0.09$  &  $-5.$  &  $-1.8$  & 
   $-1.65$  &  $-2.91$  &  $-7.$  &  $0.55$  &  $-1.31$  &  $ -3.41$  & 
   $-0.3$  &  $-0.38$  &  $0.164$  &  $-1.131$  &  --- \\ 
$0.7$  &  $0.54$  &  $1.155$  &  $0.766$  &  $0.09$  &  $-4.84$  &  $-1.74$  & 
   $-1.51$  &  $-2.78$  &  $-7.15$  &  $0.35$  &  $ -1.27$  &  $-3.3$  & 
   $-0.29$  &  $-0.36$  &  $0.282$  &  $-1.091$  &  --- \\ 
$0.9$  &  $0.544$  &  $1.163$  &  $0.774$  &  $0.1$  &  $-4.6$  &  $-1.66$  & 
   $-1.29$  &  $-2.62$  &  $-7.43$  &  $0.04$  &  $ -1.21$  &  $-3.13$  & 
   $-0.28$  &  $-0.34$  &  $0.477$  &  $-1.032$  &  --- \\ 
$1.1$  &  $0.554$  &  $1.186$  &  $0.796$  &  $0.12$  &  $-4.28$  &  $-1.57$ 
   &  $-0.97$  &  $-2.43$  &  $-7.9$  &  $-0.37$  &  $ -1.14$  &  $-2.9$  & 
   $-0.28$  &  $-0.32$  &  $0.79$  &  $-0.966$  &  --- \\ 
$1.3$  &  $0.567$  &  $1.218$  &  $0.831$  &  $0.15$  &  $-3.88$  &  $-1.46$ 
   &  $-0.48$  &  $-2.19$  &  $-8.7$  &  $-0.9$  &  $ -1.05$  &  $-2.61$  & 
   $-0.27$  &  $-0.29$  &  $1.318$  &  $-0.891$  &  --- \\ 
$1.5$  &  $0.57$  &  $1.236$  &  $0.867$  &  $0.19$  &  $-3.44$  &  $-1.34$  & 
   $0.23$  &  $-1.95$  &  $-10.24$  &  $-1.61$  &  $ -0.95$  &  $-2.28$  & 
   $-0.28$  &  $-0.26$  &  $2.363$  &  $-0.799$  &  --- \\ 
$1.7$  &  $0.529$  &  $1.138$  &  $0.786$  &  $0.17$  &  $-3.22$  &  $-1.26$ 
   &  $0.78$  &  $-2.$  &  $-14.43$  &  $-2.79$  &  $ -0.94$  &  $-2.07$  & 
   $-0.33$  &  $-0.25$  &  $5.272$  &  $-0.667$  &  --- \\
\end{tabular}
\end{table*}
\endgroup

For the $B\to X_s\gamma$ shape variables defined in Eq.~(\ref{Tdef}),
$T_i^{(1)}, \ldots T_i^{(14)}$ are independent of $E_0$, once $m_B/2-E_0 \gg
\lqcd$, and are given in the \MSbar\ scheme by
\beqa\label{msbsg1}
&& T_1^{(1)} = 2.1\,\GeV, \qquad
  T_1^{(2)} = -\frac12, \qquad
  T_1^{(3)} = T_1^{(4)} = 0, \qquad
  T_1^{(5)} = -0.06, \qquad
  T_1^{(6)} = -0.18, \nn\\*
&& T_1^{(7)} = -0.01, \qquad
  T_1^{(8)} = -0.04, \qquad
  T_1^{(9)} = -0.02, \qquad
  T_1^{(10)} = 0.37, \nn\\*
&& T_1^{(11)} = T_1^{(13)} = -0.01, \qquad
  T_1^{(12)} = T_1^{(14)} = -0.04,
\eeqa
and
\beqa\label{msbsg2}
&& T_2^{(1)} = T_2^{(2)} = T_2^{(3)} = T_2^{(4)} = 
  T_2^{(6)} = T_2^{(7)} = T_2^{(8)} = T_2^{(13)} = T_2^{(14)} = 0, \\*
&&T_2^{(5)} = -\frac1{12}, \qquad
  T_2^{(9)} = -0.04, \qquad
  T_2^{(10)} = -T_2^{(12)} = 0.06, \qquad
  T_2^{(11)} = -0.02. \nn
\eeqa
The remaining, $E_0$-dependent coefficients of the perturbative corrections are
listed in Table~\ref{tab:msbsg}.  Since in this case we are expanding the $b$
quark mass about $4.2\,$GeV, we are only showing results for $E_0 \leq
2\,$GeV.  The large size of the perturbative corrections to $T_1$ (compared to
its values in the $1S$ or PS schemes) occur to try to compensate for the bad
choice of mass scheme.

\begingroup\squeezetable
\begin{table*}[t]
\caption{Perturbative coefficients for $T_1$ and $T_2$ in the \MSbar\
scheme as a function of $E_0$.\label{tab:msbsg}}
\begin{tabular}{c||ccc||ccc}
$E_0$  &  $T_1^{(15)}$  &  $T_1^{(16)}$  &  $T_1^{(17)}$  
  &  $T_2^{(15)}$  &  $T_2^{(16)}$  &  $T_2^{(17)}$  \\ \hline\hline
$1.7$  &  $0.143$  &  $0.083$  &  $-0.009$  
  &  $0.008$  &  $0.006$  &  $-0.014$ \\ 
$1.8$  &  $0.151$  &  $0.0888$  &  $-0.002$
  &  $0.005$  &  $0.004$  &  $-0.013$ \\ 
$1.9$  &  $0.161$  &  $0.095$  &  $0.008$
  &  $0.003$  &  $0.003$  &  $-0.011$ \\ 
$2$  &  $0.173$  &  $0.106$  &  $0.03$  
  &  $0.001$  &  $0.001$  &  $-0.008$ \\ 
\end{tabular}
\end{table*}
\endgroup

\subsection{The pole mass scheme}

The $B\to X_c\ell\bar\nu$ decay width in the pole scheme is given by
\beqa\label{polewidth}
\Gamma(B\to X_c\ell\bar\nu) &=& {G_F^2\, |V_{cb}|^2\over 192\pi^3}\,
  \mBbar^5\, \bigg[ 0.370
  -0.115\, \Lambda -0.012\, \Lambda^2 + 0.\, \Lambda^3 \nn\\*
&& -0.04\, \l1 -0.10\, \l2 
  -0.01\, \l1 \Lambda + 0.02\, \l2 \Lambda \nn\\*
&& -0.02\, \r1 + 0.02\, \r2 
  -0.02\, \t1 + 0.\, \t2 \nn\\*
&& -0.03\, \t3 -0.02\, \t4 -0.040\, \epsilon
  -0.022\, \epsBLM + 0.007\, \epsilon \Lambda \bigg]\,,
\eeqa

We tabulate the shape variables defined in Eq.~(\ref{Rdef}) in
Tables~\ref{tab:poleR0}, \ref{tab:poleR1}, and \ref{tab:poleR2}, and those
defined in Eq.~(\ref{Sdef}) in Tables~\ref{tab:poleS1} and \ref{tab:poleS2} in
the pole mass scheme.  For $S_1$ and $S_2$ we do not show the $E_0$-dependence
of the order $\epsilon\Lambda$ terms, as they are not known.   For all
quantities the coefficients of the expansions are defined as in
Eq.~(\ref{expdef}).

\begingroup\squeezetable
\begin{table*}[t]
\caption{Coefficients for $R_0(0,E_0)$ in the pole scheme as a function of
$E_0$.\label{tab:poleR0}}
\begin{tabular}{c||ccccccccccccccccc}
$E_0$  &  $R_0^{(1)}$  &  $R_0^{(2)}$  &  $R_0^{(3)}$  &  $R_0^{(4)}$  &  
  $R_0^{(5)}$  &  $R_0^{(6)}$  &  $R_0^{(7)}$  &  $R_0^{(8)}$  &  
  $R_0^{(9)}$  &  $R_0^{(10)}$  &  $R_0^{(11)}$  &  $R_0^{(12)}$  &  
  $R_0^{(13)}$  &  $R_0^{(14)}$  &  $R_0^{(15)}$  &  $R_0^{(16)}$  &  
  $R_0^{(17)}$  \\ \hline\hline
$0.5$  &  $0.973$  &  $-0.002$  &  $-0.001$  &  $0.$  &  $0.$  &  $-0.01$  & 
   $0.$  &  $0.$  &  $0.$  &  $0.$  &  $0.$  &  $0.$  &  $0.$  &  $0.$  &  $
   0.$  &  $0.$  &  $0.$ \\ 
$0.7$  &  $0.93$  &  $-0.004$  &  $-0.002$  &  $0.$  &  $-0.01$  &  $-0.02$  & 
   $0.$  &  $-0.01$  &  $ 0.$  &  $0.$  &  $0.$  &  $0.$  &  $0.$  &  $0.$  & 
   $0.001$  &  $0.$  &  $0.$ \\ 
$0.9$  &  $0.86$  &  $-0.009$  &  $-0.004$  &  $0.$  &  $-0.01$  &  $-0.04$  & 
   $-0.01$  &  $-0.02$  &  $0.$  &  $0.01$  &  $ -0.01$  &  $0.$  &  $-0.01$ 
   &  $-0.01$  &  $0.001$  &  $0.$  &  $0.$ \\ 
$1.1$  &  $0.761$  &  $-0.016$  &  $-0.007$  &  $0.$  &  $-0.02$  &  $-0.08$ 
   &  $-0.02$  &  $-0.03$  &  $-0.01$  &  $0.01$  &  $ -0.01$  &  $0.$  & 
   $-0.01$  &  $-0.01$  &  $0.001$  &  $0.$  &  $0.$ \\ 
$1.3$  &  $0.634$  &  $-0.025$  &  $-0.01$  &  $0.$  &  $-0.03$  &  $-0.11$  & 
   $-0.02$  &  $-0.05$  &  $-0.01$  &  $0.02$  &  $ -0.01$  &  $0.$  & 
   $-0.02$  &  $-0.02$  &  $0.001$  &  $-0.001$  &  $0.$ \\ 
$1.5$  &  $0.483$  &  $-0.038$  &  $-0.014$  &  $-0.01$  &  $-0.05$  & 
   $-0.15$  &  $-0.03$  &  $-0.07$  &  $-0.01$  &  $ 0.02$  &  $-0.02$  & 
   $0.$  &  $-0.03$  &  $-0.03$  &  $0.$  &  $-0.002$  &  $-0.001$ \\ 
$1.7$  &  $0.318$  &  $-0.054$  &  $-0.018$  &  $-0.01$  &  $-0.06$  & 
   $-0.19$  &  $-0.04$  &  $-0.08$  &  $-0.01$  &  $ 0.02$  &  $-0.02$  & 
   $-0.01$  &  $-0.03$  &  $-0.03$  &  $-0.001$  &  $-0.003$  &  $-0.002$ \\
\end{tabular}
\end{table*}
\endgroup

\begingroup\squeezetable
\begin{table*}[t]
\caption{Coefficients for $R_1(E_0)$ in the pole scheme as a function of
$E_0$.\label{tab:poleR1}}
\begin{tabular}{c||ccccccccccccccccc}
$E_0$  &  $R_1^{(1)}$  &  $R_1^{(2)}$  &  $R_1^{(3)}$  &  $R_1^{(4)}$  &  
  $R_1^{(5)}$  &  $R_1^{(6)}$  &  $R_1^{(7)}$  &  $R_1^{(8)}$  &  
  $R_1^{(9)}$  &  $R_1^{(10)}$  &  $R_1^{(11)}$  &  $R_1^{(12)}$  &  
  $R_1^{(13)}$  &  $R_1^{(14)}$  &  $R_1^{(15)}$  &  $R_1^{(16)}$  &  
  $R_1^{(17)}$  \\ \hline\hline
$0$  &  $1.429$  &  $-0.054$  &  $-0.014$  &  $0.$  &  $-0.07$  &  $-0.18$  & 
   $-0.04$  &  $-0.05$  &  $-0.03$  &  $0.01$  &  $ -0.02$  &  $-0.01$  & 
   $-0.03$  &  $-0.03$  &  $0.$  &  $-0.002$  &  $-0.001$ \\ 
$0.5$  &  $1.459$  &  $-0.054$  &  $-0.014$  &  $0.$  &  $-0.07$  &  $-0.18$ 
   &  $-0.03$  &  $-0.05$  &  $-0.03$  &  $0.01$  &  $ -0.02$  &  $-0.01$  & 
   $-0.03$  &  $-0.03$  &  $0.$  &  $-0.002$  &  $-0.001$ \\ 
$0.7$  &  $1.498$  &  $-0.054$  &  $-0.013$  &  $0.$  &  $-0.07$  &  $-0.17$ 
   &  $-0.03$  &  $-0.05$  &  $-0.03$  &  $0.01$  &  $ -0.02$  &  $-0.02$  & 
   $-0.03$  &  $-0.03$  &  $-0.001$  &  $-0.003$  &  $-0.001$ \\ 
$0.9$  &  $1.554$  &  $-0.054$  &  $-0.013$  &  $0.$  &  $-0.07$  &  $-0.17$ 
   &  $-0.03$  &  $-0.05$  &  $-0.04$  &  $0.01$  &  $ -0.02$  &  $-0.02$  & 
   $-0.03$  &  $-0.03$  &  $-0.001$  &  $-0.003$  &  $-0.001$ \\ 
$1.1$  &  $1.625$  &  $-0.055$  &  $-0.012$  &  $0.$  &  $-0.07$  &  $-0.16$ 
   &  $-0.03$  &  $-0.05$  &  $-0.04$  &  $0.$  &  $ -0.02$  &  $-0.02$  & 
   $-0.02$  &  $-0.03$  &  $-0.002$  &  $-0.003$  &  $-0.001$ \\ 
$1.3$  &  $1.71$  &  $-0.056$  &  $-0.012$  &  $0.$  &  $-0.08$  &  $-0.15$  & 
   $-0.03$  &  $-0.05$  &  $-0.04$  &  $0.$  &  $ -0.02$  &  $-0.02$  & 
   $-0.02$  &  $-0.03$  &  $-0.002$  &  $-0.003$  &  $-0.001$ \\ 
$1.5$  &  $1.806$  &  $-0.058$  &  $-0.013$  &  $0.$  &  $-0.08$  &  $-0.14$ 
   &  $-0.04$  &  $-0.05$  &  $-0.05$  &  $-0.01$  &  $ -0.02$  &  $-0.03$  & 
   $-0.02$  &  $-0.03$  &  $-0.002$  &  $-0.003$  &  $-0.001$ \\ 
$1.7$  &  $1.913$  &  $-0.06$  &  $-0.013$  &  $0.$  &  $-0.1$  &  $-0.13$  & 
   $-0.05$  &  $-0.04$  &  $-0.07$  &  $-0.02$  &  $ -0.02$  &  $-0.04$  & 
   $-0.02$  &  $-0.02$  &  $-0.003$  &  $-0.003$  &  $-0.001$ \\
\end{tabular}
\end{table*}
\endgroup

\begingroup\squeezetable
\begin{table*}[t]
\caption{Coefficients for $R_2(E_0)$ in the pole scheme as a function of
$E_0$.\label{tab:poleR2}}
\begin{tabular}{c||ccccccccccccccccc}
$E_0$  &  $R_2^{(1)}$  &  $R_2^{(2)}$  &  $R_2^{(3)}$  &  $R_2^{(4)}$  &  
  $R_2^{(5)}$  &  $R_2^{(6)}$  &  $R_2^{(7)}$  &  $R_2^{(8)}$  &  
  $R_2^{(9)}$  &  $R_2^{(10)}$  &  $R_2^{(11)}$  &  $R_2^{(12)}$  &  
  $R_2^{(13)}$  &  $R_2^{(14)}$  &  $R_2^{(15)}$  &  $R_2^{(16)}$  &  
  $R_2^{(17)}$  \\ \hline\hline
$0$  &  $2.241$  &  $-0.184$  &  $-0.041$  &  $-0.01$  &  $-0.26$  &  $-0.58$ 
   &  $-0.11$  &  $-0.15$  &  $-0.13$  &  $0.02$  &  $ -0.08$  &  $-0.06$  & 
   $-0.09$  &  $-0.11$  &  $-0.002$  &  $-0.008$  &  $-0.003$ \\ 
$0.5$  &  $2.299$  &  $-0.185$  &  $-0.041$  &  $-0.01$  &  $-0.26$  & 
   $-0.58$  &  $-0.11$  &  $-0.15$  &  $-0.14$  &  $ 0.02$  &  $-0.08$  & 
   $-0.06$  &  $-0.09$  &  $-0.11$  &  $-0.003$  &  $-0.009$  &  $-0.003$ \\ 
$0.7$  &  $2.388$  &  $-0.188$  &  $-0.04$  &  $-0.01$  &  $-0.26$  &  $-0.57$ 
   &  $-0.11$  &  $-0.15$  &  $-0.14$  &  $0.02$  &  $ -0.08$  &  $-0.07$  & 
   $-0.09$  &  $-0.11$  &  $-0.004$  &  $-0.01$  &  $-0.003$ \\ 
$0.9$  &  $2.529$  &  $-0.193$  &  $-0.04$  &  $-0.01$  &  $-0.26$  &  $-0.57$ 
   &  $-0.11$  &  $-0.15$  &  $-0.15$  &  $0.01$  &  $ -0.08$  &  $-0.07$  & 
   $-0.09$  &  $-0.11$  &  $-0.005$  &  $-0.011$  &  $-0.003$ \\ 
$1.1$  &  $2.726$  &  $-0.2$  &  $-0.04$  &  $-0.01$  &  $-0.27$  &  $-0.56$ 
   &  $-0.11$  &  $-0.15$  &  $-0.16$  &  $0.$  &  $ -0.08$  &  $-0.08$  & 
   $-0.09$  &  $-0.11$  &  $-0.007$  &  $-0.012$  &  $-0.003$ \\ 
$1.3$  &  $2.981$  &  $-0.211$  &  $-0.041$  &  $-0.01$  &  $-0.29$  & 
   $-0.55$  &  $-0.12$  &  $-0.15$  &  $-0.18$  &  $ -0.02$  &  $-0.08$  & 
   $-0.1$  &  $-0.08$  &  $-0.1$  &  $-0.008$  &  $-0.012$  &  $-0.003$ \\ 
$1.5$  &  $3.298$  &  $-0.225$  &  $-0.043$  &  $-0.01$  &  $-0.33$  & 
   $-0.54$  &  $-0.14$  &  $-0.15$  &  $-0.22$  &  $ -0.04$  &  $-0.09$  & 
   $-0.12$  &  $-0.08$  &  $-0.1$  &  $-0.01$  &  $-0.013$  &  $-0.003$ \\ 
$1.7$  &  $3.678$  &  $-0.243$  &  $-0.047$  &  $-0.01$  &  $-0.4$  &  $-0.52$ 
   &  $-0.19$  &  $-0.14$  &  $-0.32$  &  $ -0.09$  &  $-0.1$  &  $-0.17$  & 
   $-0.07$  &  $-0.1$  &  $-0.01$  &  $-0.013$  &  $-0.003$ \\
\end{tabular}
\end{table*}
\endgroup

\begingroup\squeezetable
\begin{table*}[t]
\caption{Coefficients for $S_1(E_0)$ in the pole scheme as a function of
$E_0$.\label{tab:poleS1}}
\begin{tabular}{c||ccccccccccccccccc}
$E_0$  &  $S_1^{(1)}$  &  $S_1^{(2)}$  &  $S_1^{(3)}$  &  $S_1^{(4)}$  &  
  $S_1^{(5)}$  &  $S_1^{(6)}$  &  $S_1^{(7)}$  &  $S_1^{(8)}$  &  
  $S_1^{(9)}$  &  $S_1^{(10)}$  &  $S_1^{(11)}$  &  $S_1^{(12)}$  &  
  $S_1^{(13)}$  &  $S_1^{(14)}$  &  $S_1^{(15)}$  &  $S_1^{(16)}$  &  
  $S_1^{(17)}$  \\ \hline\hline
$0$  &  $0$  &  $1.248$  &  $0.262$  &  $0.06$  &  $1.02$  &  $-0.32$  & 
   $0.41$  &  $-0.11$  &  $0.42$  &  $-0.21$  &  $0.3$  &  $ 0.15$  &  $0.28$ 
   &  $0.08$  &  $0.102$  &  $0.111$  &  $0.038$ \\ 
$0.5$  &  $0$  &  $1.231$  &  $0.258$  &  $0.06$  &  $1.03$  &  $-0.29$  & 
   $0.41$  &  $-0.09$  &  $0.43$  &  $-0.21$  &  $0.31$  &  $ 0.16$  &  $0.28$ 
   &  $0.08$  &  $0.097$  &  $0.107$  &  --- \\ 
$0.7$  &  $0$  &  $1.209$  &  $0.251$  &  $0.06$  &  $1.05$  &  $-0.25$  & 
   $0.42$  &  $-0.07$  &  $0.45$  &  $-0.21$  &  $0.31$  &  $ 0.17$  &  $0.28$ 
   &  $0.09$  &  $0.092$  &  $0.102$  &  --- \\ 
$0.9$  &  $0$  &  $1.18$  &  $0.241$  &  $0.06$  &  $1.08$  &  $-0.18$  & 
   $0.44$  &  $-0.03$  &  $0.48$  &  $-0.2$  &  $0.31$  &  $ 0.18$  &  $0.29$ 
   &  $0.1$  &  $0.084$  &  $0.095$  &  --- \\ 
$1.1$  &  $0$  &  $1.148$  &  $0.228$  &  $0.05$  &  $1.13$  &  $-0.08$  & 
   $0.46$  &  $0.03$  &  $0.54$  &  $-0.19$  &  $0.32$  &  $ 0.21$  &  $0.29$ 
   &  $0.12$  &  $0.075$  &  $0.086$  &  --- \\ 
$1.3$  &  $0$  &  $1.118$  &  $0.211$  &  $0.04$  &  $1.22$  &  $0.04$  & 
   $0.51$  &  $0.12$  &  $0.63$  &  $-0.16$  &  $0.34$  &  $ 0.26$  &  $0.3$ 
   &  $0.14$  &  $0.064$  &  $0.077$  &  --- \\ 
$1.5$  &  $0$  &  $1.1$  &  $0.194$  &  $0.04$  &  $1.38$  &  $0.2$  &  $0.6$ 
   &  $0.26$  &  $0.8$  &  $-0.11$  &  $0.37$  &  $0.35$  &  $ 0.31$  & 
   $0.17$  &  $0.054$  &  $0.067$  &  --- \\ 
$1.7$  &  $0$  &  $1.112$  &  $0.188$  &  $0.03$  &  $1.7$  &  $0.4$  & 
   $0.83$  &  $0.47$  &  $1.16$  &  $0.04$  &  $0.43$  &  $0.53$  &  $ 0.32$ 
   &  $0.21$  &  $0.044$  &  $0.057$  &  --- \\
\end{tabular}
\end{table*}
\endgroup

\begingroup\squeezetable
\begin{table*}[t]
\caption{Coefficients for $S_2(E_0)$ in the pole scheme as a function of
$E_0$.\label{tab:poleS2}}
\begin{tabular}{c||ccccccccccccccccc}
$E_0$  &  $S_2^{(1)}$  &  $S_2^{(2)}$  &  $S_2^{(3)}$  &  $S_2^{(4)}$  &  
  $S_2^{(5)}$  &  $S_2^{(6)}$  &  $S_2^{(7)}$  &  $S_2^{(8)}$  &  
  $S_2^{(9)}$  &  $S_2^{(10)}$  &  $S_2^{(11)}$  &  $S_2^{(12)}$  &  
  $S_2^{(13)}$  &  $S_2^{(14)}$  &  $S_2^{(15)}$  &  $S_2^{(16)}$  &  
  $S_2^{(17)}$  \\ \hline\hline
$0$  &  $0$  &  $0$  &  $0.297$  &  $0.1$  &  $-3.9$  &  $0$  &  $-0.86$  & 
   $-0.82$  &  $-4.52$  &  $1.24$  &  $-0.73$  &  $-2.2$  &  $0$  &  $0$  & 
   $0.301$  &  $0.255$  &  $0.146$ \\ 
$0.5$  &  $0$  &  $0$  &  $0.294$  &  $0.1$  &  $-3.84$  &  $0$  &  $-0.83$ 
   &  $-0.8$  &  $-4.57$  &  $1.16$  &  $-0.72$  &  $ -2.17$  &  $0$  &  $0$ 
   &  $0.273$  &  $0.235$  &  --- \\ 
$0.7$  &  $0$  &  $0$  &  $0.293$  &  $0.1$  &  $-3.77$  &  $0$  &  $-0.78$ 
   &  $-0.79$  &  $-4.66$  &  $1.04$  &  $-0.71$  &  $ -2.13$  &  $0$  & 
   $0$  &  $0.241$  &  $0.212$  &  --- \\ 
$0.9$  &  $0$  &  $0$  &  $0.296$  &  $0.1$  &  $-3.67$  &  $0$  &  $-0.71$ 
   &  $-0.77$  &  $-4.83$  &  $0.87$  &  $-0.69$  &  $ -2.07$  &  $0$  & 
   $0$  &  $0.202$  &  $0.182$  &  --- \\ 
$1.1$  &  $0$  &  $0$  &  $0.301$  &  $0.1$  &  $-3.56$  &  $0$  &  $-0.61$ 
   &  $-0.75$  &  $-5.1$  &  $0.65$  &  $-0.67$  &  $ -2.01$  &  $0$  &  $0$ 
   &  $0.16$  &  $0.149$  &  --- \\ 
$1.3$  &  $0$  &  $0$  &  $0.307$  &  $0.11$  &  $-3.46$  &  $0$  & 
   $-0.48$  &  $-0.72$  &  $-5.56$  &  $0.37$  &  $-0.65$  &  $ -1.95$  & 
   $0$  &  $0$  &  $0.12$  &  $0.115$  &  --- \\ 
$1.5$  &  $0$  &  $0$  &  $0.306$  &  $0.11$  &  $-3.4$  &  $0$  &  $-0.34$ 
   &  $-0.69$  &  $-6.39$  &  $0.02$  &  $-0.64$  &  $ -1.92$  &  $0$  & 
   $0$  &  $0.083$  &  $0.084$  &  --- \\ 
$1.7$  &  $0$  &  $0$  &  $0.287$  &  $0.1$  &  $-3.43$  &  $0$  &  $-0.27$ 
   &  $-0.62$  &  $-8.05$  &  $-0.43$  &  $-0.65$  &  $ -1.94$  &  $0$  & 
   $0$  &  $0.051$  &  $0.056$  &  --- \\
\end{tabular}
\end{table*}
\endgroup

For the $B\to X_s\gamma$ shape variables defined in Eq.~(\ref{Tdef}), 
$T_i^{(1)}, \ldots T_i^{(14)}$ are independent of $E_0$, once $m_B/2-E_0 \gg
\lqcd$, and are given in the pole mass scheme by
\beqa\label{polebsg1}
&& T_1^{(1)} = \frac{\mBbar}2, \qquad
  T_1^{(2)} = -\frac12, \qquad
  T_1^{(3)} = T_1^{(4)} = T_1^{(5)} = T_1^{(7)} = 0,  \nn\\*
&& T_1^{(6)} = -0.14, \qquad
  T_1^{(8)} = -0.03, \qquad
  T_1^{(9)} = -0.02, \qquad
  T_1^{(10)} = 0.11, \nn\\*
&& T_1^{(11)} = T_1^{(13)} = 0., \qquad
  T_1^{(12)} = T_1^{(14)} = -0.03,
\eeqa
and
\beqa\label{polebsg2}
&& T_2^{(1)} = T_2^{(2)} = T_2^{(3)} = T_2^{(4)} = 
  T_2^{(6)} = T_2^{(7)} = T_2^{(8)} = T_2^{(13)} = T_2^{(14)} = 0, \\*
&&T_2^{(5)} = -\frac1{12}, \qquad
  T_2^{(9)} = -0.03, \qquad
  T_2^{(10)} = -T_2^{(12)} = 0.05, \qquad
  T_2^{(11)} = -0.02. \nn
\eeqa
The remaining, $E_0$-dependent coefficients of the perturbative corrections are
listed in Table~\ref{tab:polebsg}.

\begingroup\squeezetable
\begin{table*}[t]
\caption{Perturbative coefficients for $T_1$ and $T_2$ in the pole
scheme as a function of $E_0$.\label{tab:polebsg}}
\begin{tabular}{c||ccc||ccc}
$E_0$  &  $T_1^{(15)}$  &  $T_1^{(16)}$  &  $T_1^{(17)}$  
  &  $T_2^{(15)}$  &  $T_2^{(16)}$  &  $T_2^{(17)}$  \\ \hline\hline
$1.7$  &  $-0.077$  &  $-0.069$  &  $0.008$
  &  $0.022$  &  $0.014$  &  $-0.007$ \\ 
$1.8$  &  $-0.074$  &  $-0.068$  &  $0.012$
  &  $0.02$  &  $0.013$  &  $-0.009$ \\ 
$1.9$  &  $-0.071$  &  $-0.067$  &  $0.016$
  &  $0.017$  &  $0.012$  &  $-0.011$ \\ 
$2$  &  $-0.068$  &  $-0.065$  &  $0.021$
  &  $0.015$  &  $0.011$  &  $-0.012$ \\ 
$2.1$  &  $-0.063$  &  $-0.062$  &  $0.026$
  &  $0.012$  &  $0.009$  &  $-0.013$ \\ 
$2.2$  &  $-0.058$  &  $-0.059$  &  $0.031$
  &  $0.009$  &  $0.008$  &  $-0.013$ \\
\end{tabular}
\end{table*}
\endgroup

\end{document}